\def\({ \left( }
\def\){ \right) }
\def\b{\begin{equation}}
\def\e{\end{equation}}
\def\={\ =\ }

\def\+{\ +\ }
\def\-{\ -\ }

\def\Ls{\cal L \rm}

\def\mumu{$\mu^+\mu^-$}
\def\ee{$e^+e^-$}

\documentstyle[epsfig]{aipproc} 
\renewcommand{\thesection}{\arabic{section}}
\renewcommand{\thesubsection}{\thesection.\arabic{subsection}}

\begin{document}
\title{MUON COLLIDER DESIGN}

\author{R.~Palmer\thanks{Brookhaven National Laboratory,  Upton, NY 
11973-5000, USA}$^,$\thanks{Stanford Linear Accelerator Center, Stanford, CA 
94309, USA}, A.~Sessler\thanks{Lawrence Berkeley National Laboratory,
Berkeley, CA 94720, USA}, A.~Skrinsky\thanks{BINP, RU-630090 Novosibirsk,
Russia}, A.~Tollestrup\thanks{Fermi National Accelerator Laboratory,  
Batavia, IL 60510, USA},\\                
A.~Baltz$^1$, S.~Caspi$^{3}$, P.~Chen$^2$, W-H.~Cheng$^3$,
Y.~Cho\thanks{Argonne National Laboratory, Argonne, IL 60439-4815, USA}, 
D.~Cline\thanks{Center for Advanced Accelerators, UCLA,  Los Angeles, CA
90024-1547,USA}, E.~Courant$^1$, R.~Fernow$^1$,  J.~Gallardo$^1$, 
A.~Garren$^{3,7}$, H.~Gordon$^{1}$, M.~Green$^{3}$, R.~Gupta$^{1}$, 
A.~Hershcovitch$^1$, C.~Johnstone$^{5}$, S.~Kahn$^1$, H.~Kirk$^1$,
T.~Kycia$^1$,  Y.~Lee$^1$,  D.~Lissauer$^{1}$, A.~Luccio$^{1}$,
A.~McInturff$^{3}$,  F.~Mills$^5$, N.~Mokhov$^{5}$, G.~Morgan$^1$,
D.~Neuffer$^{5}$, K-Y.~Ng$^{5}$,
R.~Noble$^{5}$,  J.~Norem$^{6}$, B.~Norum\thanks{ University of  Virginia,
Charlottesville, VA 22901, USA}, K.~Oide \thanks{KEK, Tsukuba-shi, Ibaraki-Ken
305, Japan}, Z.~Parsa$^1$, V.~Polychronakos$^{1}$, M.~Popovic$^{5}$,
P.~Rehak$^{1}$, T.~Roser$^{1}$, R.~Rossmanith\thanks{ DESY, Hamburg, Germany}, 
R.~Scanlan$^{3}$,  L.~Schachinger$^3$, G.~Silvestrov$^4$, I.~Stumer$^1$,
D.~Summers\thanks{ University of Mississippi, Oxford, MS 38677, USA}, 
M.~Syphers$^1$,  H.~Takahashi$^1$, Y.~Torun$^{1,}$\thanks{SUNY, Stony Brook, NY  11974, USA},
 D.~Trbojevic$^1$,  W.~Turner$^{3}$,
A.~Van~Ginneken$^{5}$, T.~Vsevolozhskaya$^4$, R.~Weggel\thanks{Francis Bitter
National Magnet Laboratory, MIT, Cambridge, MA 02139, USA}, E.~Willen$^1$,
W.~Willis$^{1,}$\thanks{ Columbia University, New York, NY
10027, USA}, D.~Winn\thanks{ Fairfield University, Fairfield, CT 06430-5195,
USA}, J.~Wurtele\thanks{UC Berkeley, Berkeley, CA 94720-7300, USA}, Y.~Zhao$^1$ 
}
\maketitle
\thispagestyle{empty}
\newpage
\begin{abstract}
Muon Colliders have unique technical and physics advantages and disadvantages
when compared with both hadron and electron machines. They should thus be
regarded as complementary. Parameters are given of 4 TeV and 0.5 TeV high
luminosity \mumu colliders, and of a 0.5 TeV lower luminosity demonstration
machine.
We discuss the various systems in such muon colliders,
starting from the proton accelerator needed to generate the muons and
proceeding through muon cooling, acceleration and storage in a collider ring.
Detector background, polarization, and nonstandard operating conditions are 
discussed.
\end{abstract}

\section{INTRODUCTION}
\subsection{Technical Considerations}
The possibility of muon colliders was introduced  by Skrinsky et
al.\cite{ref2} and Neuffer\cite{ref3}. More recently, several
workshops and collaboration meetings have greatly increased the level of
discussion\cite{ref4},\cite{ref5}.  In this paper we present scenarios for
4 TeV and 0.5 TeV  colliders based on an optimally designed proton source, 
and
for a lower luminosity 0.5 TeV demonstration based on an upgraded version of
the AGS. It is assumed that a demonstration version based on upgrades of the
FERMILAB machines would also be possible (see second Ref.\cite{ref5}).

Hadron collider energies are limited by their size, and technical constraints
on bending magnetic fields. At very high energies it will also become
impractical to obtain the required luminosities, which must rise as the energy
squared. \ee colliders, because they undergo simple, single-particle
interactions, can reach higher energy final states than an equivalent hadron
machine. However, extension of $e^+e^-$  colliders to multi-TeV energies is
severely performance-constrained by beamstrahlung, and cost-constrained because
two full energy linacs are required\cite{ref1} to avoid the excessive
synchrotron radiation that would occur in rings. Muons (${m_{\mu}\over
m_e}=207$) have the same advantage in energy reach as electrons, but have
negligible beamstrahlung, and can be accelerated and stored in rings, making
the possibility of high energy \mumu colliders attractive. There are however, 
several major technical problems with $\mu$'s:

\begin{itemize}
\item they decay with a lifetime of  $2.2\times 10^{-6}$ s. This problem is
partially overcome by rapidly increasing the energy of the muons, and thus
benefiting from their
relativistic $\gamma$ factor. At 2 TeV, for example, their lifetime is
$0.044\,$s: sufficient for approximately 1000 storage-ring collisions;
 \item another consequence of the muon decays
is that the decay products heat the magnets of the collider
ring and  create backgrounds in the detector;

\item  Since the muons are created through pion decay into a diffuse phase
space, some form of cooling is essential. Conventional stochastic or
synchrotron cooling is too slow to be effective before they decay. Ionization
cooling, can be used, but the final emittance of the muon beams will remain
larger than that  possible for electrons in an \ee collider. 
\end{itemize}

Despite these problems it appears possible that high energy muon colliders 
might have luminosities comparable to or, at energies of several TeV, even 
higher than those in \ee colliders\cite{ref6}. And because the \mumu machines
would be much  smaller\cite{ref7}, and require much lower precision (the final
spots are about  three orders of magnitude larger), they may be significantly
less expensive.  It must be remembered, however, that a \mumu collider remains
a new and  untried concept, and its study has just begun; it cannot yet be
compared with  the more mature designs for an \ee collider.

\subsection{Physics Considerations}
There are at least two physics advantages of a \mumu collider, when compared
with an \ee collider:
   \begin{itemize}
\item Because of the lack of beamstrahlung, a \mumu collider can be operated
with an energy spread of as little as 0.01 \%. It is thus possible to use the
\mumu collider for precision measurements of masses and widths, that would be
hard, if not impossible, with an \ee collider.
\item The direct coupling of a lepton-lepton system to a Higgs boson has a
cross section that is proportional to the square of the mass of the lepton. As
a result, the cross section for direct Higgs production from the \mumu system
is 40,000 times that from an \ee system. 
\end{itemize}

However, there are liabilities:
   \begin{itemize}
   \item It will be relatively hard to obtain both high polarization and good 
luminosity in a \mumu collider, whereas good polarization of one beam can be 
obtained in an \ee collider without any loss in luminosity. One notes however 
that in the muon case, moderate polarization could be obtained for both 
beams. 
   \item because of the decays of the muons, there will be a considerable
background of photons, muons and neutrons in the detector. This background may
be acceptable for some experiments, but it cannot be as clean as in an 
\ee collider.
   \end{itemize}
\subsection{Discussion}
   We conclude that a muon collider has both technical advantages and 
disadvantages when compared with an \ee machine. Similarly it has specific 
physics advantages and disadvantages. It thus seems reasonable to consider 
\mumu colliders as complementary to \ee colliders, just as \ee colliders are 
complementary to hadron machines. 
\subsection{Overview of Components}
\begin{figure}[t!] 
\centerline{\epsfig{file=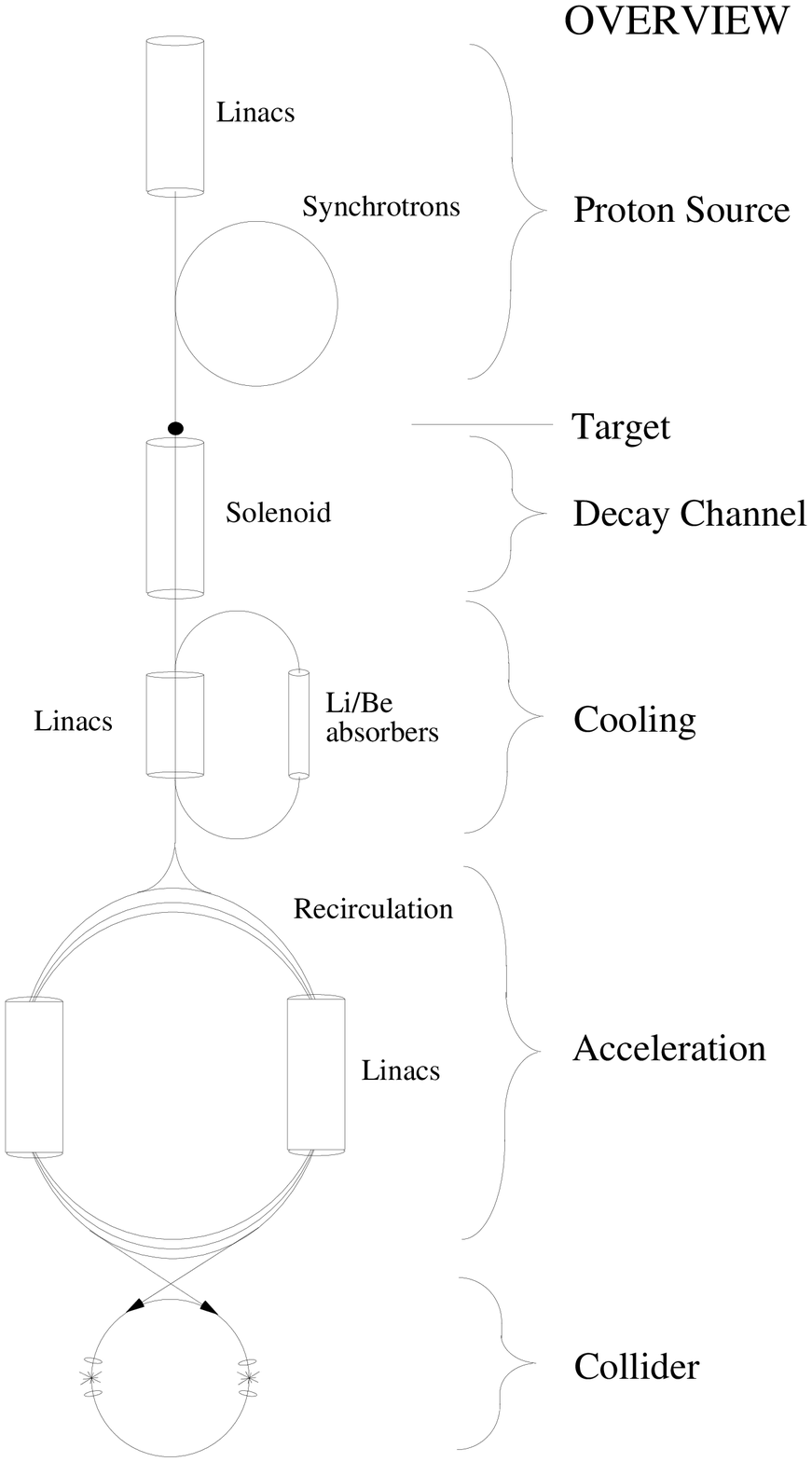,height=4.0in,width=3.5in}}
\caption{Schematic of a Muon Collider.}
\label{schematic}
\end{figure}

The basic components of the \mumu collider are shown schematically in 
Fig.\ref{schematic}. Tb.\ref{sum} shows parameters for the candidate 
designs. The normalized  emittance $\epsilon^N$ is defined as the {\it rms}
transverse phase space divided by $\pi .$ Notice that more precisely a factor
of $\pi $ must appear in the dimensions of emittance (i.e. $\pi\,{\rm mm\,mrad}
 $).  
   A high intensity proton source  is bunch compressed and focussed on a heavy 
metal target. The pions generated are captured by a high field solenoid and 
transferred to a solenoidal decay channel within a low frequency linac. The 
linac serves to reduce, by phase rotation, the momentum spread of the pions, 
and of the muons into which they decay. Subsequently, the muons are cooled by 
a sequence of ionization cooling stages. Each stage consists of energy loss, 
acceleration, and emittance exchange by energy absorbing wedges in the 
presence of dispersion. Once they are cooled the muons must be rapidly 
accelerated to avoid decay. This can be done in recirculating accelerators 
(\`{a} la CEBAF) or in fast pulsed synchrotrons. Collisions occur in a separate 
high field collider storage ring with a single very low beta insertion. 
\begin{table}[thb]  
\begin{tabular}{llccc}
                           &          & 4 TeV & .5 TeV & Demo.  \\
\tableline
Beam energy                & TeV      &     2    &   .25  &  .25  \\
Beam $\gamma$              &          &   19,000 &  2,400   &  2,400 \\
Repetition rate            & Hz       &    15    &    15    &    2.5 \\
Muons per bunch            & $10^{12}$  &   2    &    4     &    4   \\
Bunches of each sign       &          &   2      &    1     &    1   \\
Normalized {\it rms} emittance $\epsilon^N$   &$\pi$mm mrad  &  50  &  90   &   90   \\
Bending Field              &  T   &    9    &    9    &    8   \\
Circumference              &  Km      &    7    &    1.2  &   1.5   \\
Average ring mag. field $B$    & T   & 6    &   5     &    4  \\
Effective turns before decay &       & 900    &   800  &   750 \\
$\beta^*$ at intersection   & mm     &   3   &   8   &   8    \\
{\it rms} beam size at I.P.       & $\mu m$&   2.8 &  17   &   17   \\
Luminosity &${\rm cm}^{-2}{\rm s}^{-1}$& $10^{35}$&5\,$10^{33}$& $6\,10^{32}$\\
\end{tabular}
\caption{Parameters of Collider Rings}
\label{sum}
\end{table}

\section{MUON PRODUCTION}

\subsection{Proton Driver}

The specifications of the proton drivers are given in Tb.\ref{driver}. In the
examples, it is a high-intensity ($2.5\times  10^{13}$ protons per pulse) 30
GeV proton synchrotron. The preferred cycling rate would be 15 Hz, but for a
demonstration machine using the AGS\cite{roser}, the repetition rate would be
limited to 2.5 Hz and to $24\,$GeV. For the lower energy machines, 2 final
bunches are employed (one to make $\mu^-$'s and the other to make $\mu^+$'s).
For the high energy collider, four are used  (two $\mu$ bunches of each sign).
\begin{table}[b!] 
\begin{tabular}{llccc}
                           &          & 4 TeV    & .5 TeV   &  Demo \\
\tableline
Proton energy              & GeV      &    30    &   30     &   24  \\
Repetition rate            & Hz       &    15    &    15    &    2.5 \\
Protons per bunch          & $10^{13}$&   2.5    &    2.5   &    2.5   \\
Bunches                    &          &   4      &    2     &    2   \\
Long. phase space/bunch    & eV s    &   4.5    &    4.5   &    4.5  \\
Final {\it rms} bunch length     & ns     &    1     &    1     &    1    \\
\end{tabular}
 \caption{Proton Driver Specifications}
\label{driver}
\end{table}

Earlier studies had suggested that the driver could be a 10 GeV machine with
the same charge per bunch, but a repetition rate of $30\,$Hz. This
specification was almost identical to that studied\cite{ref77} at ANL for a
spallation neutron source. Studies at FNAL\cite{mills} have further established
that such a specification is reasonable. But in order to reduce the cost
of the muon phase rotation section and for minimizing the final muon
longitudinal phase space, it appears now that the final proton bunch length
should be 1 ns (or even less). This appears difficult to achieve at 10 GeV, but
possible at 30 GeV.

A 1 ns rms bunch at 30 GeV with a phase space per bunch of
$6\pi\,\sigma_t\sigma_E=4.5\,{\rm eV s}$ (at 95\%)  bunch,  will have a
momentum spread of $0.8\,\%,$  ($2\, \%$ at 95\%), and the space charge tune
shift just  before extraction would be $\approx 0.5.$ Provided the rotation can
be performed rapidly enough, this should not be a problem.

An attractive technique \cite{compress} for bunch compression would be to 
generate a large momentum  spread with modest rf at a final energy close to
transition. Pulsed quads would then be employed to move the operating point
away from transition, resulting in rapid compression.

\subsection{Target and Pion Capture}

\begin{figure}[t!] 
\centerline{\epsfig{file=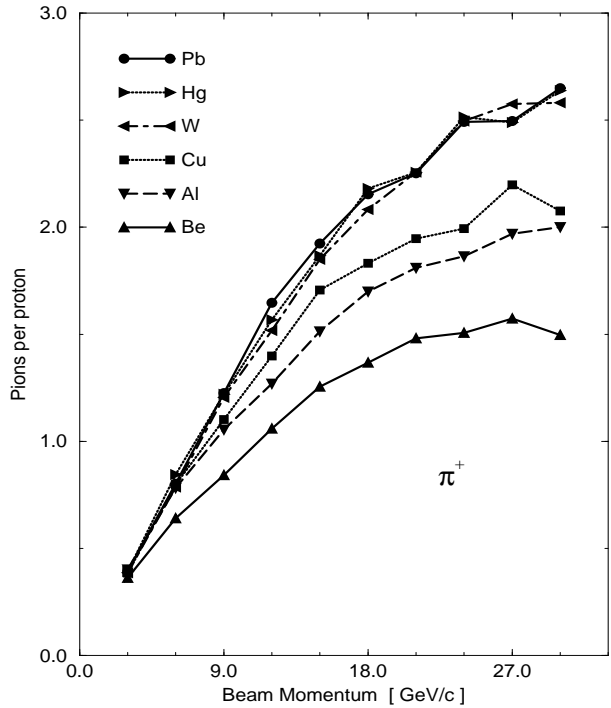,height=4.0in,width=3.5in}}
\caption{ARC forward $\pi^+$ production vs proton energy and target material.}
\label{pionproda}
\end{figure}

Predictions of the nuclear Monte-Carlo program ARC\cite{arc} suggest that $\pi$
production is maximized by the use of heavy target materials, and that the
production is peaked at a relatively low pion energy ($\approx 100\,$MeV),
substantially independent of the initial proton energy. Fig.\ref{pionproda} 
shows the forward $\pi^+$ production as a function of proton energy and target
material; the $\pi^-$ distributions are similar. 
Other programs\cite{MARS},\cite{other} do not predict such a large low energy 
peak, and there is currently very little data to indicate which is right. An 
experiment (E910), currently running at the AGS, should decide this question, 
and thus settle at which energy the capture should be optimized. 
\begin{figure}[t!] 
\centerline{\epsfig{file=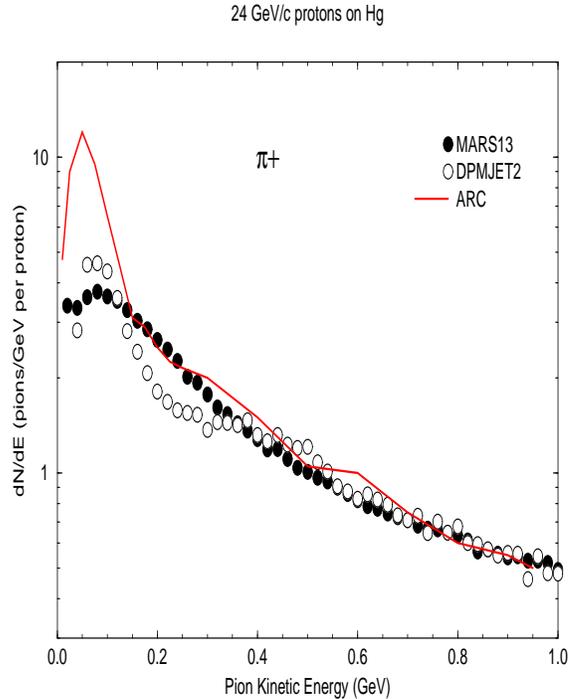,height=4.0in,width=3.5in}}
\caption{$\pi^{+}$ energy distribution for $24\,$GeV protons on Hg. }
 \label{pionprodp}
 \end{figure}
\begin{figure}[thb] 
\centerline{\epsfig{file=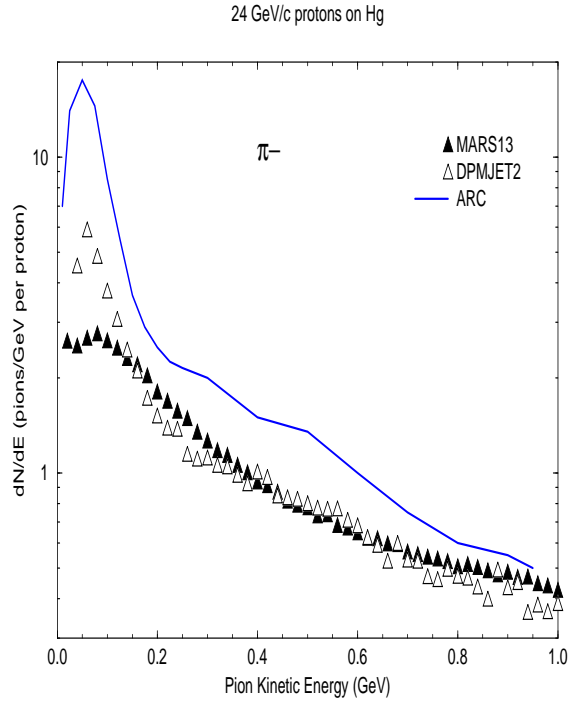,height=4.0in,width=3.5in}}
\caption{$\pi^{-}$ energy distribution for $24\,$GeV protons on Hg. }
 \label{pionprodm}
 \end{figure}
The target would probably be made of Cu, approximately 24 cm long by 2 cm 
diameter. A study\cite{snake} indicates that, with a 3 mm rms beam, the single 
pulse instantaneous temperature rise is acceptable, but, if cooling is only 
supplied from the outside, the equilibrium temperature would be excessive. 
Some method must be provided to give cooling within the target volume. For 
instance, the target could be made of a stack of relatively thin copper disks, 
with water cooling between them. 

Figs.\ref{pionprodp},\ref{pionprodm} compared the predictions of the mentioned
codes, for the energy  distribution of $\pi^+$ and $\pi^-$ for 24 GeV protons on
Hg; the  distributions for Cu are similar. 
 
\begin{figure}[b!] 
\centerline{\epsfig{file=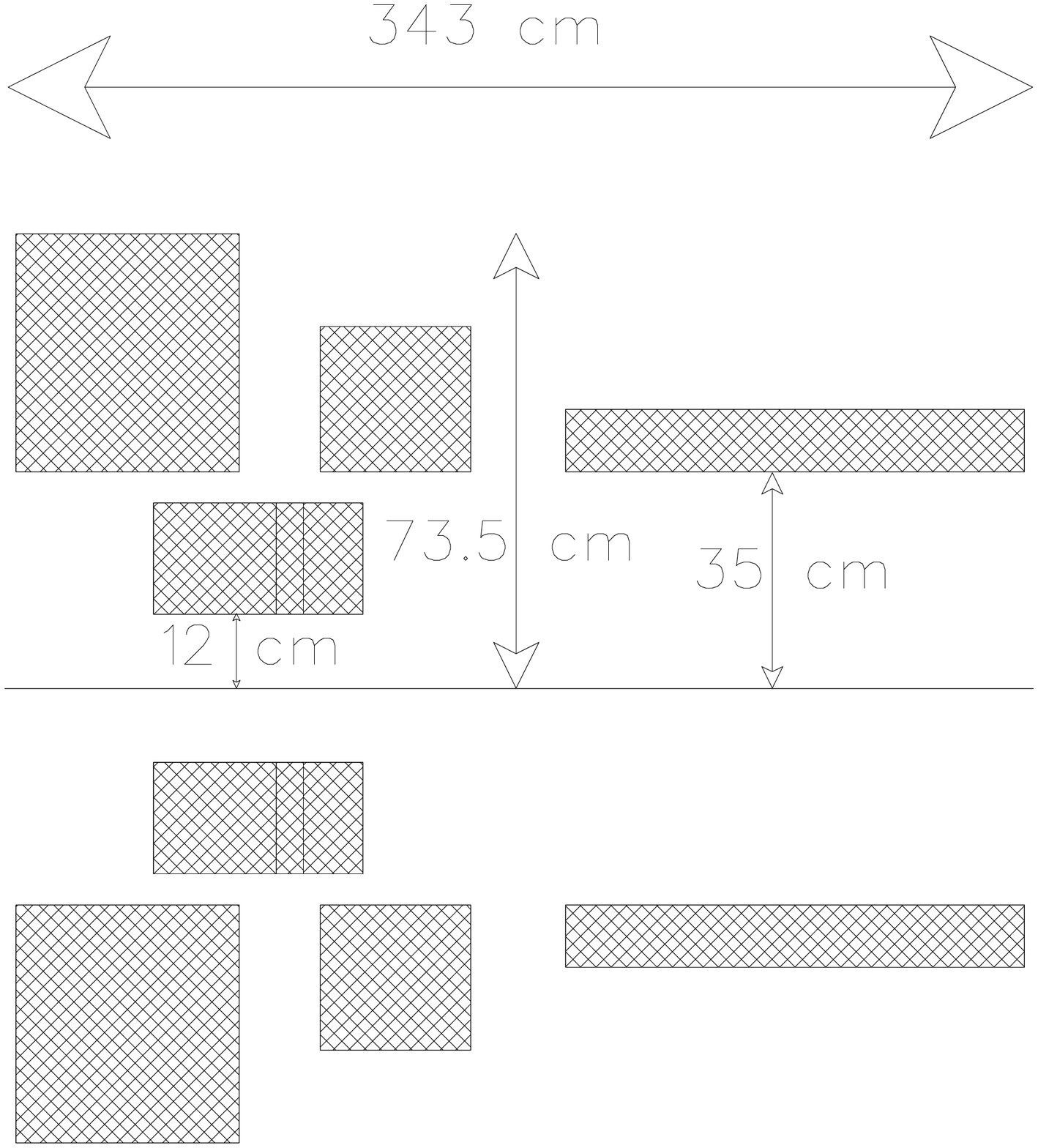,height=4.0in,width=4.0in}}
\caption{Schematic of a hybrid magnet solenoid system for $\pi$ capture and matching.}
 \label{bitter}
 \end{figure}

\begin{figure}[b!] 
\centerline{\epsfig{file=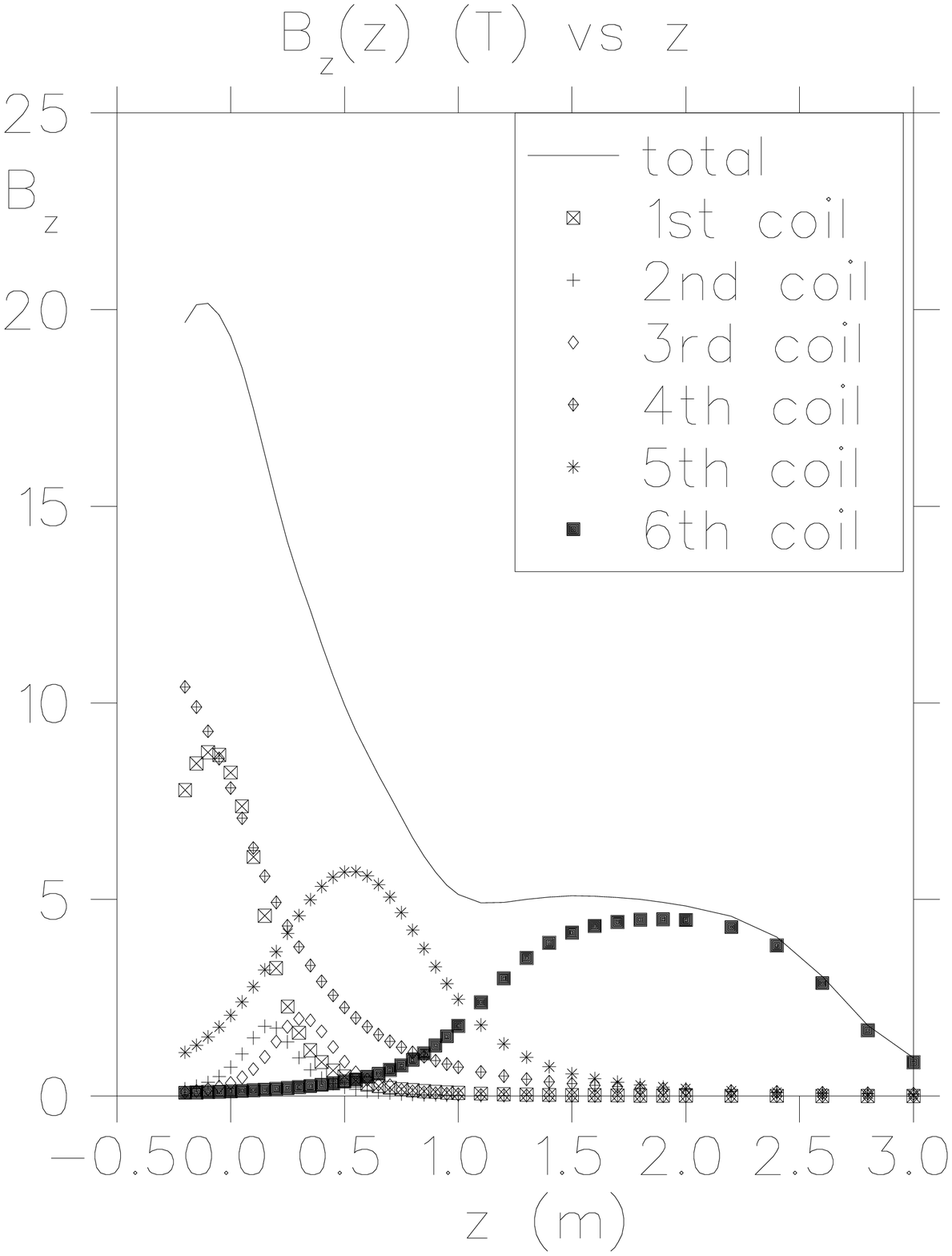,height=4.0in,width=4.0in}}

\caption{Total and individual component field profiles of hybrid magnet solenoid.}
 \label{matchB}
 \end{figure}

Pions are captured from the target by a high-field ($20\,$T, 15 cm inside 
diameter) hybrid magnet: superconducting on the outside, and a water cooled 
Bitter solenoid on the inside.   A preliminary design\cite{weggel} (see 
Fig.\ref{bitter}) has an inner Bitter magnet with an inside diameter of 24 cm 
(space is allowed for a 4 cm heavy metal shield inside the coil) and an 
outside diameter of 60 cm; it provides half (10T) of the total field, and 
would consume approximately 8 MW. The superconducting magnet has a set of 
three coils, all with inside diameters of 70 cm and is designed to give 10 T 
at the target and provide the required tapered field\cite{ref9} (see 
Fig.\ref{matchB}) to match into the periodic superconducting solenoidal decay 
channel ($5\,$T and radius $=15\,$cm). 

Monte Carlo studies indicate a yield of 0.4--0.6 muons, of each sign, per initial 
proton, captured in the decay channel. Surprisingly, this conclusion seems 
relatively independent of whether the system is optimised for energies of 50 
to 500 MeV (using ARC), or 200 to 2000 MeV (using MARS). 

\subsection{ Use of Both Signs}
Protons on the target produce pions of both signs, and a solenoid will capture 
both, but the required subsequent phase rotation rf systems will have opposite 
effects on each. One solution is to break the proton bunch into two, aim them 
on the same target one after the other, and adjust the rf phases such as to 
act correctly on one sign of the first bunch and on the other sign of the 
second. This is the solution assumed in the parameters of this paper. 

A second possibility would be to: 1) to separate the charges into two channels, 
2) delay the particles of one charge by introducing a chikane in one of the 
channels, 3) recombine the two channels so that the particles of the two 
charges are in line, but separated longitudinally (i.e. box cared). Both 
charges can now be phase rotated by a single linac with appropriate phases of 
rf. 

A third solution is to separate the pions of each charge prior to the use
of rf, and feed the beams of each charge into different channels. 

After the target, and prior to the use of any rf or cooling, the beams have 
very large emittances and energy spread. Conventional charge separation using 
a dipole is not practical. But if a solenoidal channel is bent, then the 
particles trapped within that channel will drift\cite{drift} in a direction 
perpendicular to the bend. With our parameters this drift is dominated by a 
term (curvature drift) that is linear with the forward momentum of the 
particles, and has a  direction that depends on the sign of the charges. 
If sufficient bend is employed\cite{snake}, the two 
charges could then be separated by a septum and captured into two separate 
channels. When these separate channels are bent back to the same forward 
direction, the momentum dispersion is separately removed in each new 
channel. 

Although this idea is very attractive, it has some problems:

\begin{itemize}
\item If the initial beam has a radius r=$0.15\,$m, and if the momentum range
to be accepted is $F={p_{{\rm max}}\over p_{{\rm min}}}=3,$ then the required
height of the solenoid just prior to separation is 2(1+{\it F})r=$1.2\,$m. Use of a
lesser height will result in particle loss. Typically, the reduction in yield
for a curved solenoid compared to a straight solenoid is about $25\,\%$ (due to
the loss of very low and very high momentum pions), but this must be weighed
against the fact that both charge signs are captured for each proton on target.
\item The system of bend, separate, and return bend will require significant
length and must occur prior to the start of phase rotation (see below).
Unfortunately, it appears that the cost of the phase rotation rf is 
strongly dependent on keeping this distance as short as possible. On the other
hand a bent solenoid would separate the remnant proton beam and other charged
debris exiting the target before the rf cavities. 
 \end{itemize}
Clearly, compromises will be involved, and more study of this concept is 
required.                          

\subsection{Phase Rotation Linac}

The pions, and the muons into which they decay, have an energy spread from
about 0 - 3 GeV, with an rms/mean of $\approx 100 \%$, and with a peak at
about 100 MeV. It would be difficult to handle such a wide spread in any
subsequent system. A linac is thus introduced along the decay channel, with
frequencies and phases chosen to deaccelerate the fast particles and
accelerate the slow ones; i.e. to phase rotate the muon bunch. Tb.\ref{rot}
gives an example of parameters of such a linac. It is seen that the lowest
frequency is 30 MHz, a low but not impossible frequency for a conventional
structure.

\begin{table}[thb]
\begin{tabular}{cccc}
Linac     & Length & Frequency & Gradient  \\
          &  m     &   MHz     &  MeV/m    \\
\tableline
1         &  3    &   60     &   5      \\
2         &  29   &   30     &   4       \\
3         &  5    &   60      &   4      \\
4         &  5    &   37     &   4       \\
\end{tabular}
\caption{Parameters of Phase Rotation Linacs}
\label{rot}
\end{table}

\begin{figure}[tbh] 
\centerline{\epsfig{file=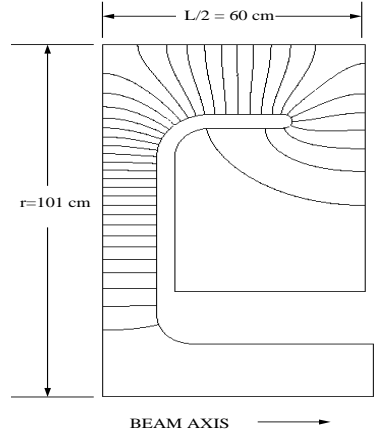,height=3.25in,width=3.25in}}
\vskip 1cm
\caption{30 MHz cavity for use in phase rotation and early stages of cooling.}
\label{30MHz}
\end{figure}

A design of a reentrant 30 MHz cavity is shown in Fig.\ref{30MHz}. Its
parameters are given in Tb.\ref{30MHzt}.
\begin{table}[thb]
\begin{tabular}{llc}
Cavity Radius    & cm & 101 \\
Cavity Length  &   cm & 120  \\
Beam Pipe Radius  &  cm  &  15  \\
Accelerating Gap  &  cm  &  24  \\
Q               &        &  18200  \\
Average Acceleration Gradient & MV/m  &  3  \\
Peak rf Power     &  MW  &  6.3  \\
Average Power (15 Hz) & KW & 18.2  \\
Stored Energy   & J  &  609  \\
\end{tabular}
\caption{Parameters of 30 MHz rf Cavity}
\label{30MHzt}
\end{table}
It has a diameter of approximately 2 m, only about one third of that of a
conventional pill-box  cavity. To keep its cost down, it would be made of Al.
Multipactoring would probably be suppressed by stray fields from the 5 T
focusing coils, but could also be controlled by an internal coating of titanium
nitride.

\begin{figure}[bht] 
\centerline{\epsfig{file=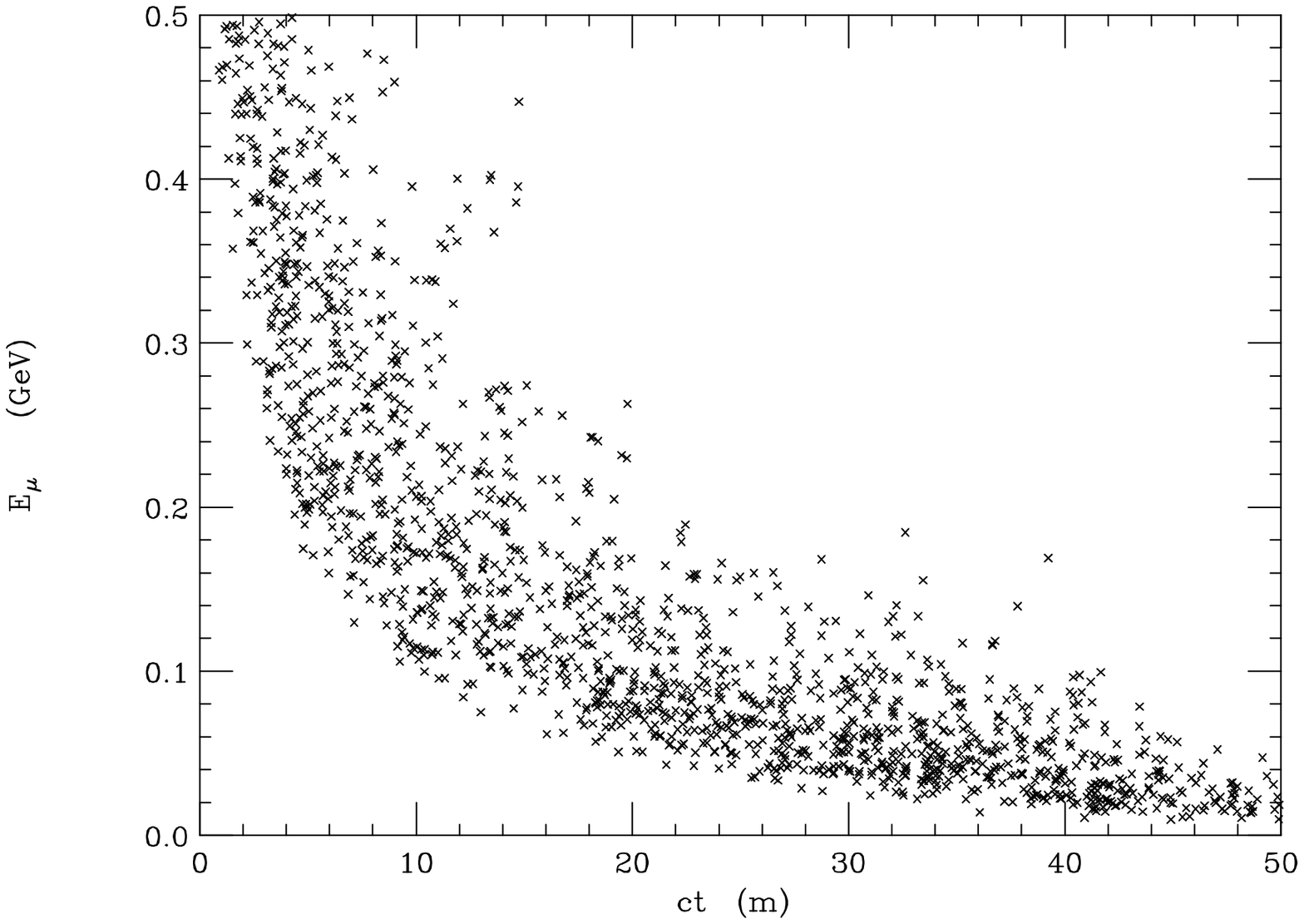,height=4.0in,width=3.5in}}
\caption{Energy vs ct of Muons at End of Decay Channel without Phase 
Rotation.}
\label{Evsct1}
 \end{figure}

\begin{figure}[bht] 
\centerline{\epsfig{file=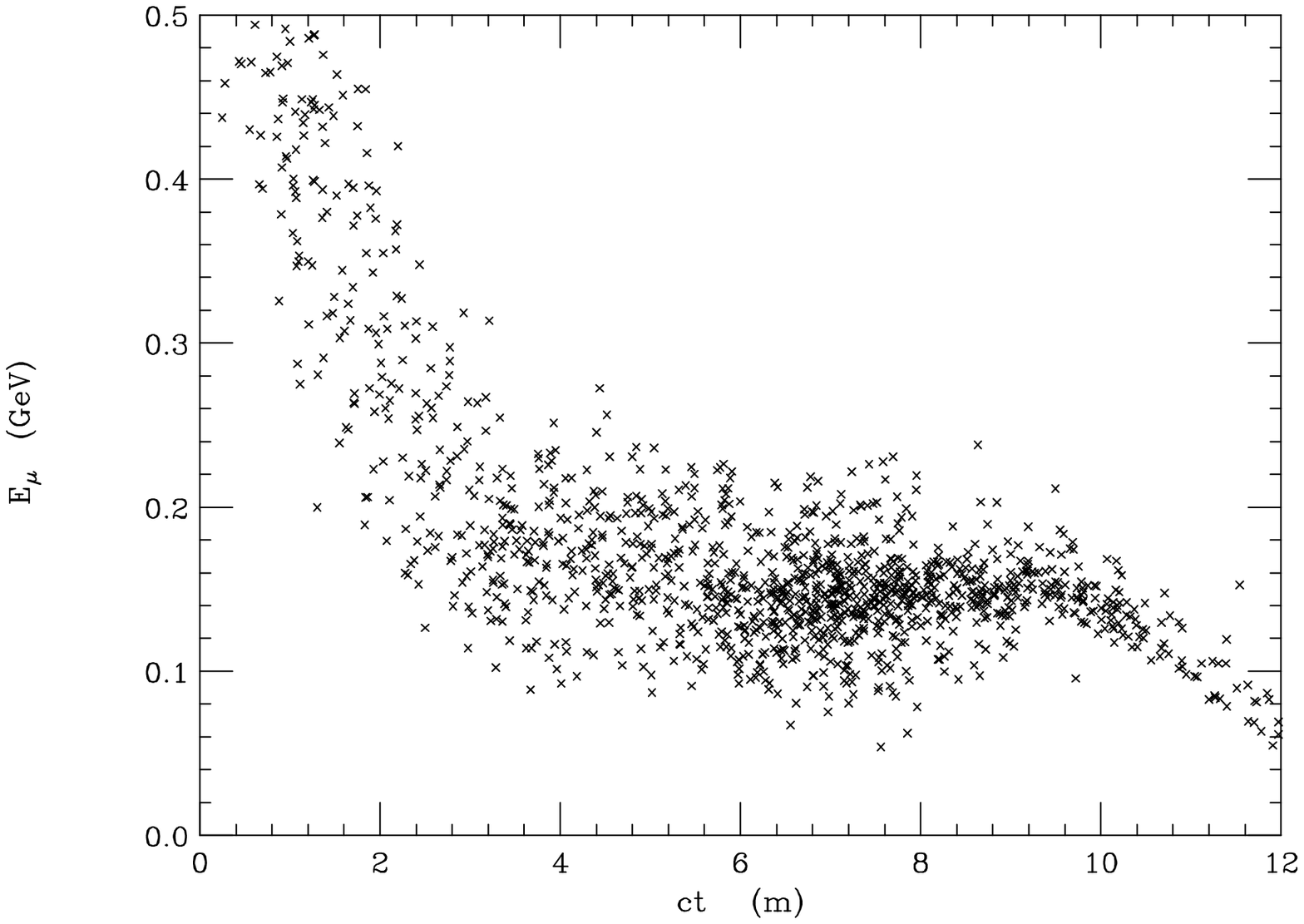,height=4.0in,width=3.5in}}
\caption{Energy vs ct of Muons at End of Decay Channel with Phase 
Rotation.} 
 \label{Evsct2}
 \end{figure}

  Figs.\ref{Evsct1} and \ref{Evsct2}  show the energy vs c$\,$t at the end 
of the decay channel with and without phase rotation. Note that the c$\,$t 
scales are very different: the rotation both compacts the energy spread and 
limits the growth of the bunch length. 

After this phase rotation, a bunch can be selected with mean energy 150 MeV,
rms bunch length $1.7\,$m, and rms momentum  spread  $20\,$\% ($95\,$\%,
$\epsilon_{\rm L}= 3.2\,{\rm eV s}$). The number of
muons per initial proton in this selected  bunch is 0.2--0.3, about half the
total number of pions initially captured.  As noted above, since the linacs
cannot phase rotate both signs in the same bunch, we need two  bunches: the
phases are set to rotate the $\mu^+$'s of one bunch and the  $\mu^-$'s of the
other. Prior to cooling, the bunch is accelerated to 300 MeV,  in order to
reduce the momentum spread to $10\,\%.$

\section{COOLING}
For collider intensities, the phase-space volume must be reduced within the
$\mu$ lifetime. Cooling by synchrotron radiation, conventional stochastic
cooling and conventional electron cooling are all too slow. Optical stochastic
cooling\cite{ref11}, electron cooling in a plasma discharge\cite{ref12} and
cooling in a crystal lattice\cite{ref13} are being studied, but appear very
difficult. Ionization cooling\cite{ref14} of muons seems relatively
straightforward.

\subsection{Ionization Cooling Theory}

In ionization cooling, the beam loses both transverse and longitudinal momentum
as it passes through a material medium. Subsequently, the longitudinal
momentum can be restored by coherent reacceleration, leaving a net loss of
transverse momentum. Ionization cooling is not practical for protons and
electrons because of nuclear interactions (p's) and bremsstrahlung (e's), 
but is practical for $\mu$'s because of their low nuclear
cross section and relatively low bremsstrahlung.

The equation for transverse cooling (with energies in GeV)  is:
  \begin{equation}
\frac{d\epsilon_n}{ds}\ =\ -\frac{dE_{\mu}}{ds}\ \frac{\epsilon_n}{E_{\mu}}\ +
\ \frac{\beta_{\perp} (0.014)^2}{2\ E_{\mu}m_{\mu}\ L_R},\label{eq1}
  \end{equation}
where $\epsilon_n$ is the normalized emittance, $\beta_{\perp}$ is the betatron
function at the absorber, $dE_{\mu}/ds$ is the energy loss, and $L_R$  is the
radiation length of the material.  The first term in this equation is the coherent
cooling term, and the second is the heating due to multiple scattering.
This heating term is minimized if $\beta_{\perp}$ is small (strong-focusing)
and $L_R$ is large (a low-Z absorber). From Eq.\ref{eq1} we find a limit to transverse cooling, which occurs when
heating due to multiple scattering  balances cooling due to energy loss. The
limits are $\epsilon_n\approx\ 0.6\ 10^{-2}\ \beta_{\perp}$ for Li, and
$\epsilon_n\approx\ 0.8\  10^{-2}\ \beta_{\perp}$ for Be.

The equation for energy spread  (longitudinal emittance) is:
 \begin{equation}
{\frac{d(\Delta E)^2}{ds}}\ =\
-2\ {\frac{d\left( {\frac{dE_\mu}{ds}} \right)} {dE_\mu}}
\ <(\Delta E_{\mu})^2 >\ +\
{\frac{d(\Delta E_{\mu})^2_{{\rm straggling}}}{ds}}\label{eq2}
 \end{equation}
where the first term is the cooling (or heating) due to energy loss, 
and the second term is the heating due to straggling.

Cooling requires that  ${d(dE_{\mu}/ds)\over dE_{\mu}} > 0.$ But at energies
below about 200 MeV, the energy loss function for muons, $dE_{\mu}/ds$, is
decreasing with energy and there is thus heating of the beam.
Above 400 MeV the energy loss function increases gently, giving some
cooling, but not sufficient for our application.

Energy spread can also be reduced by artificially increasing
${d(dE_\mu/ds)\over dE_{\mu}}$ by placing a transverse variation in absorber
density or thickness at a location where position is energy dependent, i.e. where there is
dispersion. The use of such wedges can reduce energy spread, but it
simultaneously increases transverse emittance in the direction of the
dispersion. Six dimensional phase space is not reduced, but it does allow the
exchange of emittance between the longitudinal and transverse directions.

In the long-path-length Gaussian-distribution limit, the heating term (energy
straggling)  is given by\cite{ref15}
 \begin{equation}
\frac{d(\Delta E_{\mu})^2_{{\rm straggling}}}{ds}\ =\
4\pi\ (r_em_ec^2)^2\ N_o\ \frac{Z}{A}\ \rho\gamma^2\left(1-
\frac{\beta^2}{2}\right),
 \end{equation}
where $N_o$ is Avogadro's number and $\rho$ is the density. Since the energy
straggling increases as $\gamma^2$, and the cooling system size scales as
$\gamma$, cooling at low energies is desired.

\subsection{Cooling System}

We require a reduction of the normalized transverse emittance by almost three
orders of magnitude (from $1\times 10^{-2}$ to $5\times 10^{-5}\,$m-rad), and a
reduction of the longitudinal emittance by one order of magnitude.
This cooling is obtained in a series of cooling stages. In general, each stage
consists of three components with matching sections between them:

 \begin{enumerate}
 \item a FOFO lattice consisting of spaced axial solenoids with alternating field
directions and lithium hydride
absorbers placed at the centers of the spaces between
them, where the $\beta_{\perp}$'s are minimum.
 \item a lattice consisting of more widely separated alternating solenoids,
and bending magnets between them
to generate dispersion. At the location of maximum
dispersion, wedges of lithium hydride are introduced to interchange
longitudinal and transverse emittance.
 \item a linac to restore the energy lost in the absorbers.
 \end{enumerate}

   At the end of a sequence of such cooling stages, the transverse emittance can
be reduced to about $10^{-3}\,$m-rad, still a factor of $\approx 20$ above the
emittance goals of Tb.\ref{sum}. The longitudinal emittance, however, can be
cooled to a value nearly three orders of magnitude less than  is required. The
additional reduction of transverse emittance can then be obtained by  a
reverse exchange of transverse and longitudinal phase-spaces. This is again
done by the use of wedged absorbers in dispersive regions between solenoid
elements.

   Throughout this process appropriate momentum compaction and rf fields must
be used to control the bunch, in the presence of space charge, wake field and
resistive wall effects.

In a few of the later stages, current carrying lithium rods might replace item
(1) above. In this case the rod serves simultaneously to maintain the low
$\beta_{\perp}$, and attenuate the beam momenta. Similar lithium rods, with
surface fields of $10\,$T , were developed at Novosibirsk and have been used as
focusing elements at FNAL and CERN\cite{ref16}. It is hoped\cite{20Tli} that
liquid lithium columns, can be used to raise the surface field to 20 T and
improve the resultant cooling. The Li or Be lenses will permit smaller
$\beta_{\perp}$ and therefore more transverse cooling with the consequence that
the emittance exchange with the longitudinal would be reduced. 

It would be desirable, though not necessarily practical, to economize on
linac sections by forming groups of stages into recirculating loops.

\subsection{Example}

A {\it model}
example has been generated that uses no lithium rods and no recirculating
loops. It is assumed here that each charge is cooled in a separate channel, 
although it might be possible to design a system with both charges 
in the same channel. Individual components of the lattices have been defined, but a
complete lattice has not yet been specified, and no Monte Carlo study of its
performance has yet been performed. Spherical aberration due to solenoid end
effects, wake fields, and second order rf effects have not yet been included.

The phase advance in each cell of the lattice is made as close to $\pi$ as
possible in order to minimize the $\beta$'s at the location of the absorber,
but it is kept somewhat less than this value so that the phase advance per cell
should never exceed $\pi$. The following effects are included: the maximum
space charge transverse defocusing; a $3\, \sigma$ fluctuation of momentum; a 
$3\, \sigma$ fluctuation in amplitude.

\begin{figure}[bht] 
\centerline{\epsfig{file=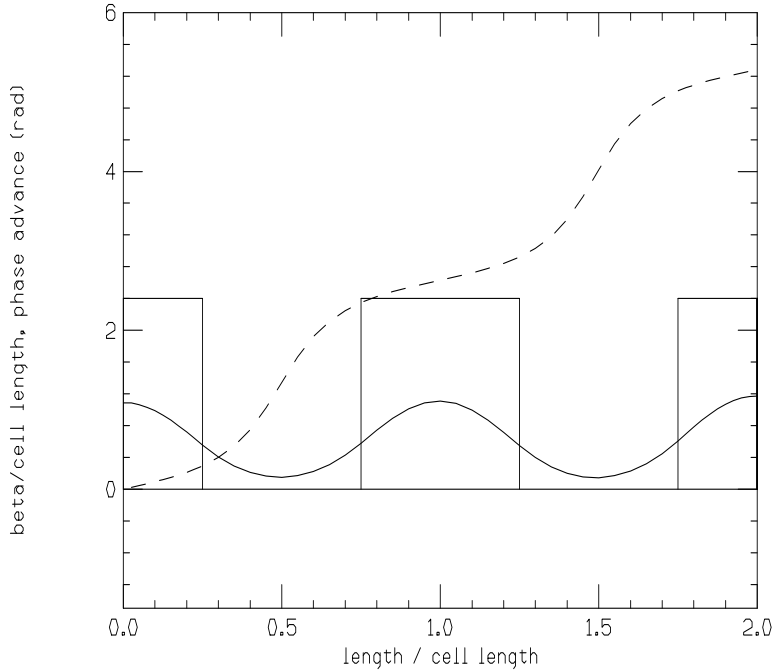,height=3.5in,width=4.0in}}
\caption{Beta function(solid) and Phase Advance(dashed) vs $z$ in FOFO cells.}
 \label{fofo}
 \end{figure}

\begin{figure}[bht] 
\centerline{\epsfig{file=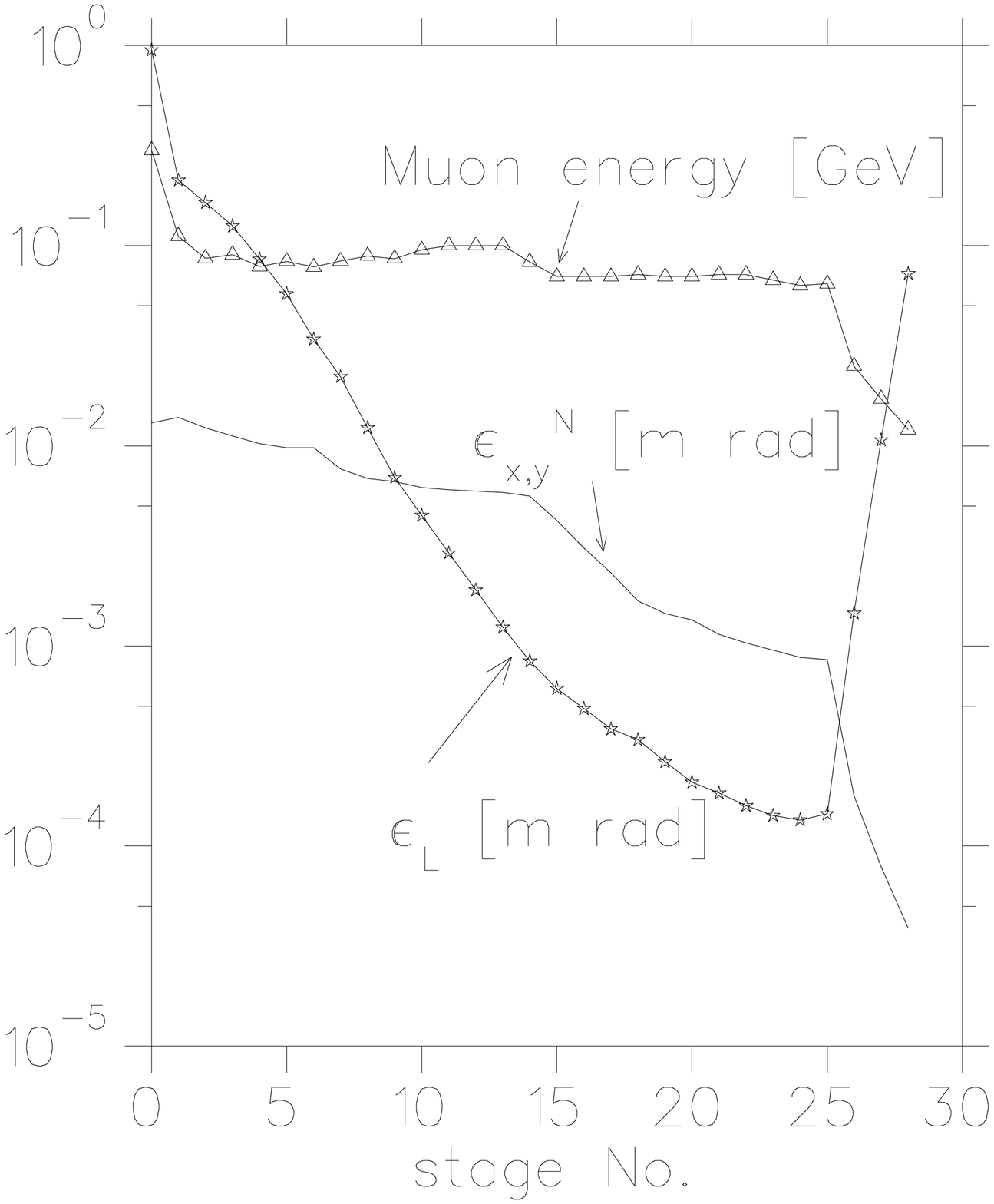,height=4.0in,width=3.5in}}
\caption{$\epsilon_{\perp}$, ${\epsilon_{L}\,c\over \left< {\rm E}_{\mu}\right>} $ and E$_{\mu}$ [GeV] vs stage number in the cooling sequence. }
 \label{cooling}
 \end{figure}

Fig.\ref{fofo} shows the beta function (solid-line) and phase advance
(dashed-line) through two  typical cells of the cooling lattice.  In the early
stages, the solenoids  have relatively large diameters and their fields are
limited to $7\,$T. In later stages the emittance has decreased, the apertures
are smaller and the fields are increased to $11\,$T. The maximum bending fields
used are $7\,$T, but most are closer to $3\,$T.
The emittances, transverse and longitudinal, as a function of stage number, are
shown in Fig.\ref{cooling}, together with the beam energy. In the first 15
stages, relatively strong wedges are used to rapidly reduce the longitudinal
emittance, while the transverse emittance is reduced relatively slowly. The
object is to reduce the bunch length, thus allowing the use of higher frequency
and higher gradient rf in the reacceleration linacs. In the next 10 stages, the
emittances are reduced close to their asymptotic limits. In the last  two 
stages, the transverse and longitudinal emittances are again exchanged, but in
the opposite direction: lowering the transverse and raising the longitudinal.
During this exchange the energy is allowed to fall to 10 MeV in order to
minimize the $\beta$, and thus limit the emittance dilution.

   The total length of the system is 500 m, and the total acceleration used is
3.3 GeV. The fraction of muons remaining at the end of the cooling system
is calculated to be $58\,$\%.

\section{ACCELERATION}
Following cooling and initial bunch compression the beams must be rapidly 
accelerated
to full energy (2 TeV, or 250 GeV). A sequence of linacs would work, but would
be expensive. Conventional synchrotrons cannot be used because the muons would
decay before reaching the required energy. The conservative solution is to use
a sequence of recirculating accelerators (similar to that used at CEBAF). A
more economical solution would be to use fast rise time pulsed magnets in
synchrotrons, or synchrotrons with rapidly rotating permanent magnets
interspersed with high field fixed magnets.

\subsection{Recirculating Acceleration}
Tb.\ref{acceleration} gives an example of a possible sequence of
recirculating accelerators. After initial linacs, there are two conventional rf recirculating accelerators taking the muons
up to 75 GeV, then two superconducting recirculators going up to 2000 GeV.
\begin{table}
 \begin{tabular}{llccccc}
                     &         & Linac &  \#1   &  \#2   &   \#3  &   \#4  \\
\tableline
initial energy       & GeV     &   0.20&     1 &     8 &    75 &   250 \\
final energy         & GeV     &     1 &     8 &    75 &   250 &  2000 \\
nloop                &         &     1 &    12 &    18 &    18 &    18 \\
freq.                & MHz     &   100 &   100 &   400 &  1300 &  2000 \\
linac V              & GV      &   0.80&   0.58&   3.72&   9.72&  97.20 \\
grad                 &         &     5 &     5 &    10 &    15 &    20  \\
dp/p  initial        & \%      &    12 &   2.70&   1.50&     1 &     1  \\
dp/p  final          & \%      &   2.70&   1.50&     1 &     1 &   0.20 \\
$\sigma_z$ initial   & mm      &   341 &   333 &  82.52&  14.52&   4.79 \\
$\sigma_z$ final     & mm      &   303 &  75.02&  13.20&   4.36&   3.00 \\
$\eta$               & \%      &   1.04&   0.95&   1.74&   3.64&   4.01 \\
$N_\mu$              & $10^{12}$ &   2.59&   2.35&   2.17&   2.09&   2    \\
$\tau_{fill}$        & $\mu$s  &  87.17&  87.17&  10.90&  s.c. &  s.c.  \\
beam t               & $\mu$s  &   0.58&   6.55&  49.25&   103 &   805   \\
decay survival       &         &   0.94&   0.91&   0.92&   0.97&   0.95  \\
linac len            & km      &   0.16&   0.12&   0.37&   0.65&   4.86  \\
arc len              & km      &   0.01&   0.05&   0.45&   1.07&   8.55  \\
tot circ             & km      &   0.17&   0.16&   0.82&   1.72&  13.41  \\
phase slip           & deg     &     0 &  38.37&   7.69&   0.50&   0.51  \\
\end{tabular}
 \caption{Parameters of Recirculating Accelerators} 
\label{acceleration}
\end{table}

\begin{figure}[t!] 
\centerline{\epsfig{file=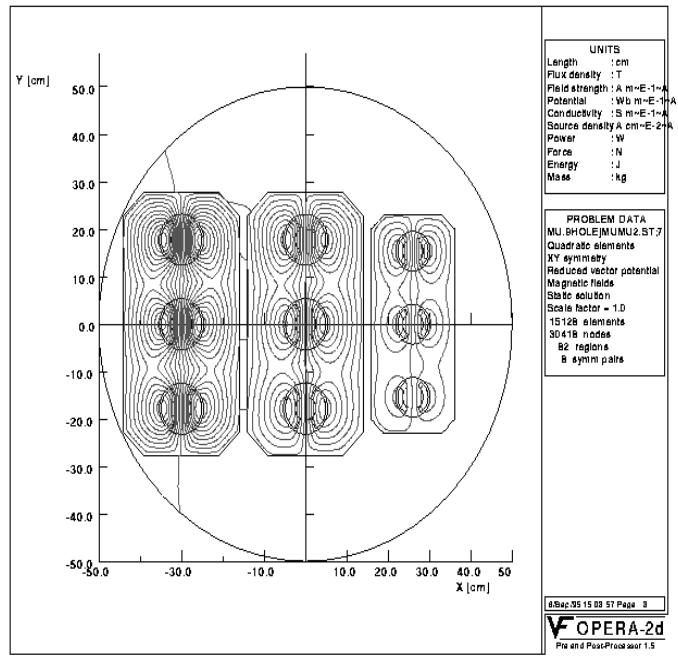,height=4.0in,width=4.5in}}
\caption{A cross section of a 9 aperture sc magnet.  }
 \label{9hole}
 \end{figure}

   Criteria that must be considered in picking the parameters of such
accelerators are:
\begin{itemize}
\item The wavelengths of rf should be chosen to limit the loading, $\eta$, (it
is restricted to below 4 \% in this example) to avoid excessive longitudinal
wakefields and the resultant emittance growth.
 \item  The wavelength should also be sufficiently large compared to the bunch
length to avoid second order effects (in this example: 10 times).
 \item  For power efficiency, the cavity fill time should be
long compared to the acceleration time. When conventional cavities cannot
satisfy this condition, superconducting cavities are required.
 \item  In order to minimize muon decay during acceleration (in this example
73\% of the muons are accelerated without decay), the number of
recirculations at each stage should be kept low, and the rf acceleration
voltage correspondingly high. But for minimum cost, the number of
recirculations appears to be of the order of 20 - a relatively high number. In
order to avoid a large number of separate magnets, multiple aperture magnets
can be designed (see Fig.\ref{9hole}).
 \end{itemize}

Note that the linacs see two bunches of opposite signs, passing through
in opposite directions. In the final accelerator in the 2 TeV case, each bunch
passes through the linac 18 times. The total loading
is then $4\times 18\times \eta = 288 \%.$ With this loading, assuming 60\%
klystron efficiencies and reasonable cryogenic loads, one could probably
achieve 35\% wall to beam power efficiency, giving a wall power
consumption for the rf in this ring of 108 MW.

A recent study\cite{neufferacc} tracked particles through a
similar sequence of recirculating accelerators and found a dilution of
longitudinal phase space of the order of 10\% and negligible particle loss.

\subsection{Pulsed Magnets}

\begin{figure}[t!] 
\centerline{\epsfig{file=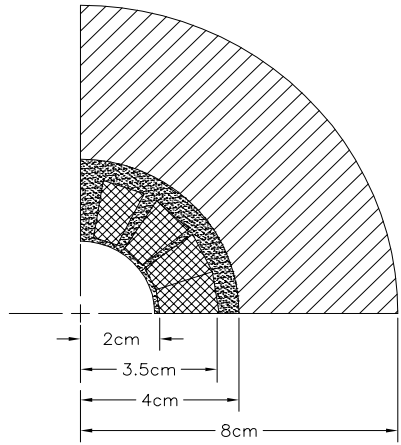,height=4.1in,width=2.9in}}
\caption{Cross section of pulsed magnet for use in the
acceleration to 250 GeV.  }
 \label{pulse}
 \end{figure}

An alternative to recirculating accelerators for stages \#2 and \#3 would be to
use pulsed magnet synchrotrons. The cross section of a pulsed magnet for this
purpose is shown in Fig.\ref{pulse}. If desired, the number of recirculations
could be higher in this case, and the needed rf voltage correspondingly lower,
but the loss of particles from decay would be somewhat more. The cost for a
pulsed magnet system appears to be significantly less than that of a multi-hole
recirculating magnet system, and the power consumption is moderate for energies 
up to 250 GeV.  Unfortunately, the power consumption is impractical at
energies above about 500 GeV.

\subsection{Pulsed and Superconducting Hybrid}

For the final acceleration to 2 TeV in the high energy machine, the power 
consumed by a ring using only pulsed magnets would be excessive. A 
recirculating accelerator is still usable, but a hybrid ring with alternating 
pulsed warm magnets and fixed superconducting magnets might be a cheaper 
alternative. 

   For instance: A sequence of two such hybrid accelerators could be used. 
One from 0.25-1 TeV, and the other from 1-2 TeV. Each would 
employ 10 T superconducting fixed magnets alternating with 
pulsed warm magnets whose fields would swing from -1.5 T to + 1.5 T. The power 
consumption of such a system would be large, but the capital cost probably 
far  less than that of a recirculating accelerator.

\subsection{Rotating and Superconducting Hybrid}
   Pulsed magnets would be used up to the highest possible energy; say 0.5 
TeV. A sequence of two hybrid accelerators could then be used: one from $0.5-
1\,{\rm TeV},$ and the other from 1-2 TeV. Each would employ 8 T 
superconducting fixed magnets alternating now with pairs of counter-rotating 
permanent magnets\cite{rotating}. Fields as high as 2 T should be possible in 
magnets of outside diameter less than 20 cm. 

If, for example, the superconducting magnets are 1.5 times the lengths of the  
two  rotating magnets, then as the rotating magnets turn, the average field 
will swing from ${(1.5 x 8-2-2) \over 3.5}=2.3$ T to ${(1.5 x 8+2+2) \over 
3.5}=4.6$; varying by a factor of 2. For the final stage (1 to 2 TeV): with 
acceleration in 20 turns, and a ring circumference of 20 km, the acceleration 
time would be 1.3 msec, requiring a rotation rate of about 15,000 rpm. For the 
penultimate stage (0.5 to 1 TeV): again with acceleration in 20 turns, and a 
ring circumference of 10 km, the acceleration time would be 0.7 msec, 
requiring a rotation rate of about 30,000 rpm. If 30,000 rpm is too high, 40 
turn acceleration could be used in this stage and the rotation rate kept at 
15,000 rpm. This should be practical, and the power consumption would be 
negligible. However, many technical questions remain to be answered. 

\section{COLLIDER STORAGE RING} After acceleration, the $\mu^+$ 
and $\mu^-$ bunches are injected into a separate storage ring. The highest 
possible average bending field is desirable, to maximize the number of 
revolutions before decay, and thus maximize the luminosity. Collisions would 
occur in one, or perhaps two, very low-$\beta^*$ interaction areas. Parameters 
of the ring were given earlier in  Tb.\ref{sum}.

\subsection{Bending Magnet Design}

   The magnet design is complicated by the fact that the $\mu$'s decay within
the rings ($\mu^-\ \rightarrow\ e^-\overline{\nu_e}\nu_{\mu}$), producing
electrons whose mean energy is approximately $0.35$ that of the muons. These
electrons travel toward the inside of the ring dipoles, radiating a 
fraction of their energy as synchrotron radiation towards the outside of the
ring, and depositing the rest on the inside.  The total average power deposited,
in the ring, in the 4 TeV machine is 13 MW, yet the maximum power that can 
reasonably be taken from the magnet coils at 4$\,$K is only of the order of 
40 KW. The power deposited could be reduced if the beams are kicked 
out of the ring prior to their their complete decay. Since the luminosity goes 
as the square of the number of muons, a significant power reduction can be 
obtained for a small luminosity loss. But still the power level is high. Two 
promising approaches are discussed below. 
\subsubsection{Large Cosine-Theta Magnet}
\begin{figure}[b!] 
\centerline{\epsfig{file=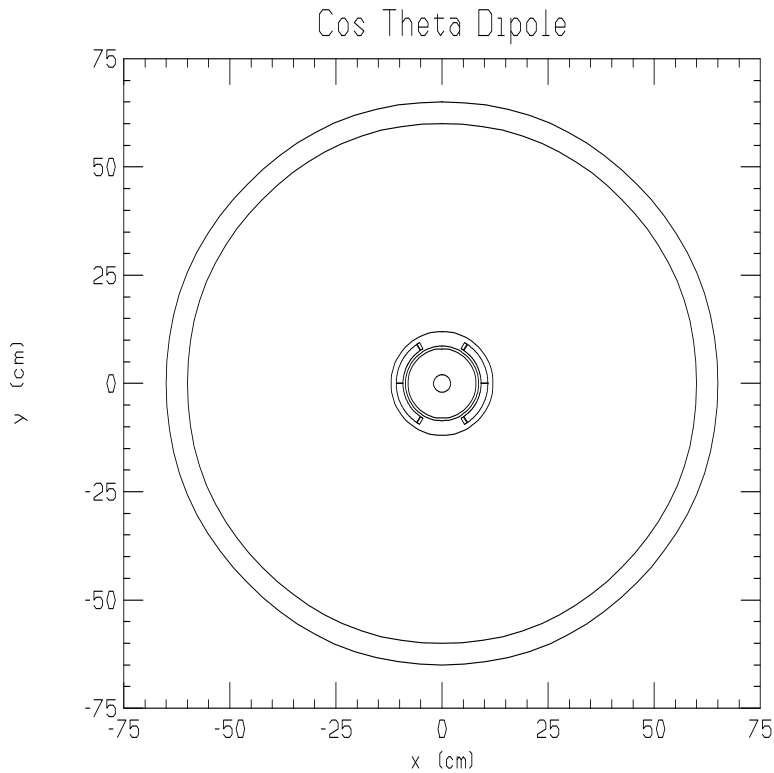,height=4.0in,width=4.0in}}
\caption{Cos Theta Arc Bending Magnet}
 \label{costhetamag}
 \end{figure}

\begin{figure}[b!] 
\centerline{\epsfig{file=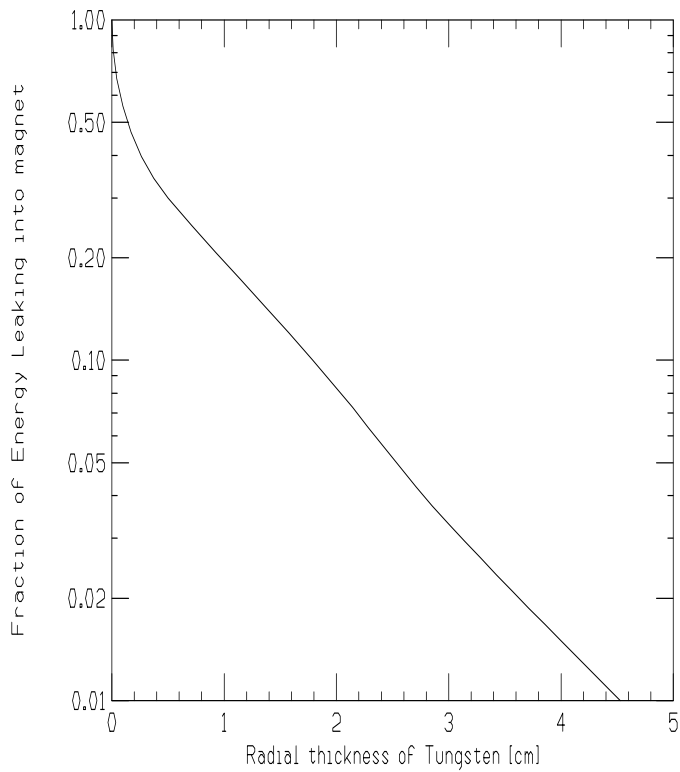,height=4.0in,width=3.5in}}
\caption{Energy attenuation vs the thickness of a tungsten liner. }
 \label{shield}
 \end{figure}

The beam is surrounded by a thick warm shield, located inside a large aperture 
conventional cosine-theta magnet (see Fig.\ref{costhetamag}). Fig.\ref{shield} 
shows the attenuation of the heating produced as a function of the thickness 
of a warm tungsten liner\cite{Iuliupipe}. If conventional superconductor is 
used, then the thicknesses required in the three cases would be as given in 
Tb.\ref{linert}. If high Tc superconductors could be used, then these 
thicknesses could probably be halved. 

If this approach were taken, then the quadrupoles would best use warm iron
poles  placed as close to the beam as practical. The coils could be either 
superconducting or warm, as dictated by cost considerations. If an elliptical 
vacuum chamber were used, and the poles were at 1 cm radius, then gradients of 
150 T/m should be possible. 

 \begin{table}  
 \begin{tabular}{llccc}
                  &        &  2TeV   &   0.5 TeV   &    Demo   \\
\tableline
Unshielded Power  &   MW   &  13     &   1.6        &    .26   \\
Liner inside rad &    cm   &  2    &    2   &   2     \\
Liner thickness  &   cm   &   6   &    4      &     2    \\
Coil inside rad  &   cm   &   9    &   7     &     5     \\ 
Attenuation      &        &   400  &   80     &   12   \\
Power leakage     &   KW   &  32   &    20      &     20   \\
Wall power for $4\,K$ &  MW   &   26   &    16     &   16    \\
\end{tabular}
\caption{Thickness of Shielding for Cos Theta Collider Magnets.}
\label{linert}
\end{table}

\subsubsection{`C' Magnets} 
\begin{figure}[b!] 
\centerline{\epsfig{file=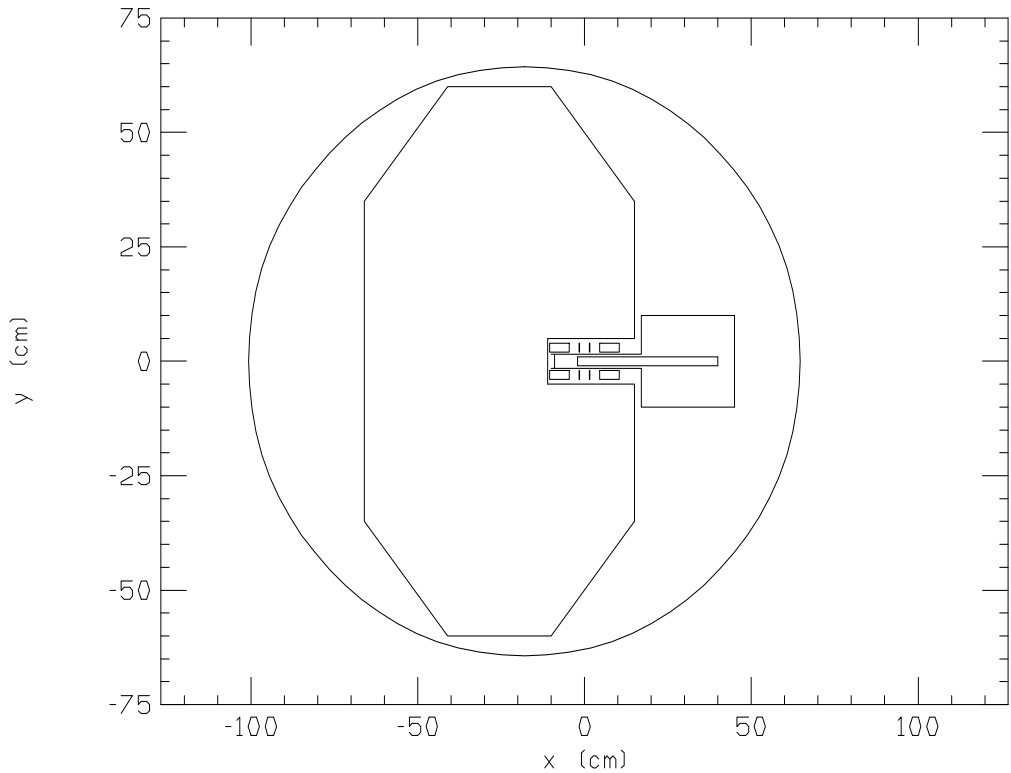,height=4.0in,width=5.25in}}
\caption{`C' Arc Bending Magnet.}
 \label{cmag} 
 \end{figure}

  An alternative is to use `C' magnets facing inward, with a broad 
vacuum pipe extending out of the gaps (see Fig.\ref{cmag}). The 
decay electrons will spiral inward, out of the gap of the magnet, and can be 
absorbed in a separate warm dump. Some of the synchrotron radiation will still 
strike the inner wall of the vacuum chamber and a more limited dump is 
required there. 

With ${\rm Nb_3 Sn}$ conductors, there appear no theoretical problems in achieving 10 
T fields with very good field quality $(dB/B \leq 10^{-5}$ for $x \leq 1 {\rm
cm})$. The 
problems would be in supporting the coils and maintaining the required position 
accuracy. 

\begin{figure}[b!] 
\centerline{\epsfig{file=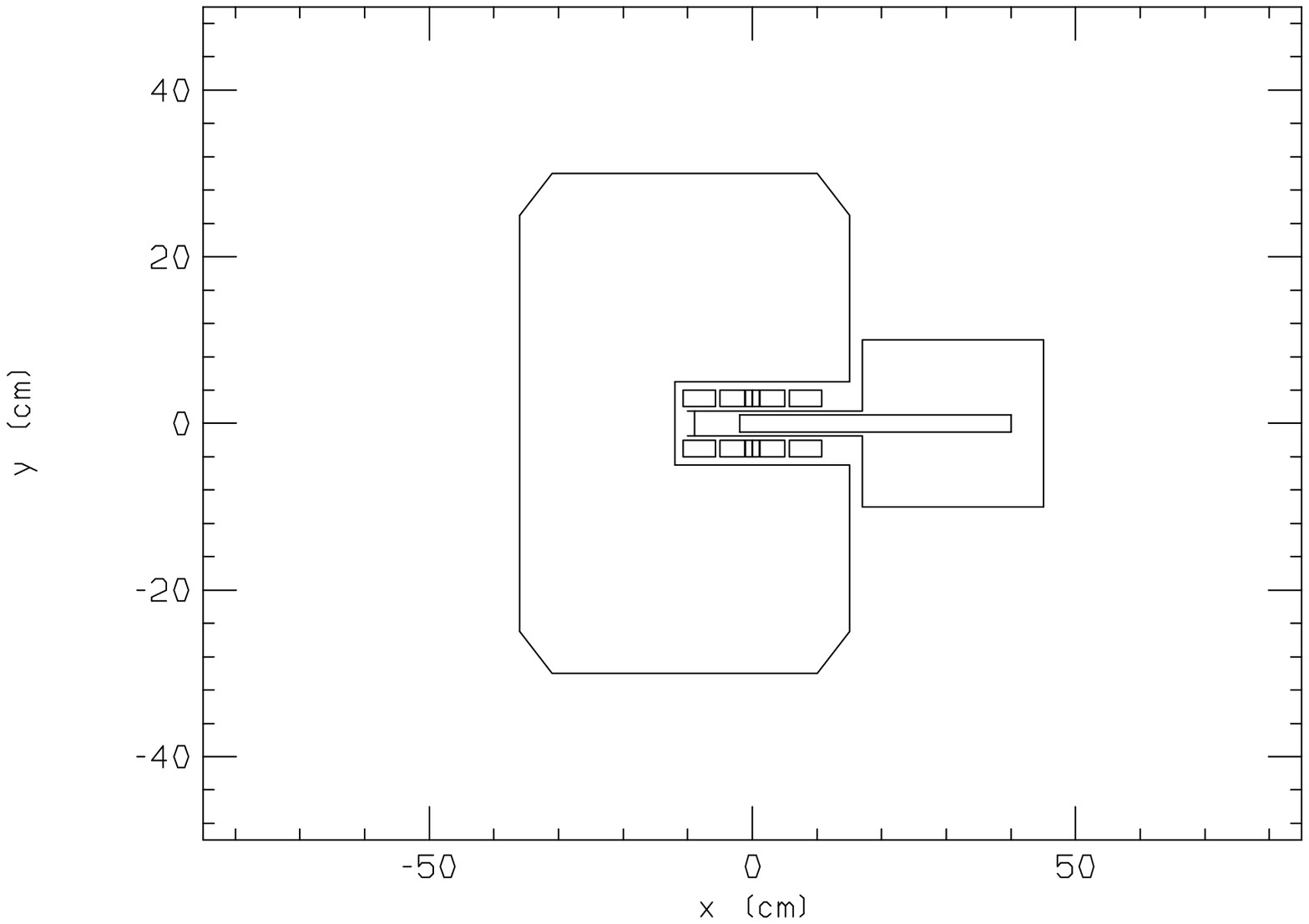,height=4.0in,width=5.5in}}
\caption{`C' Arc Quadrupole.}
 \label{cquad}
 \end{figure}

\begin{figure}[b!] 
\centerline{\epsfig{file=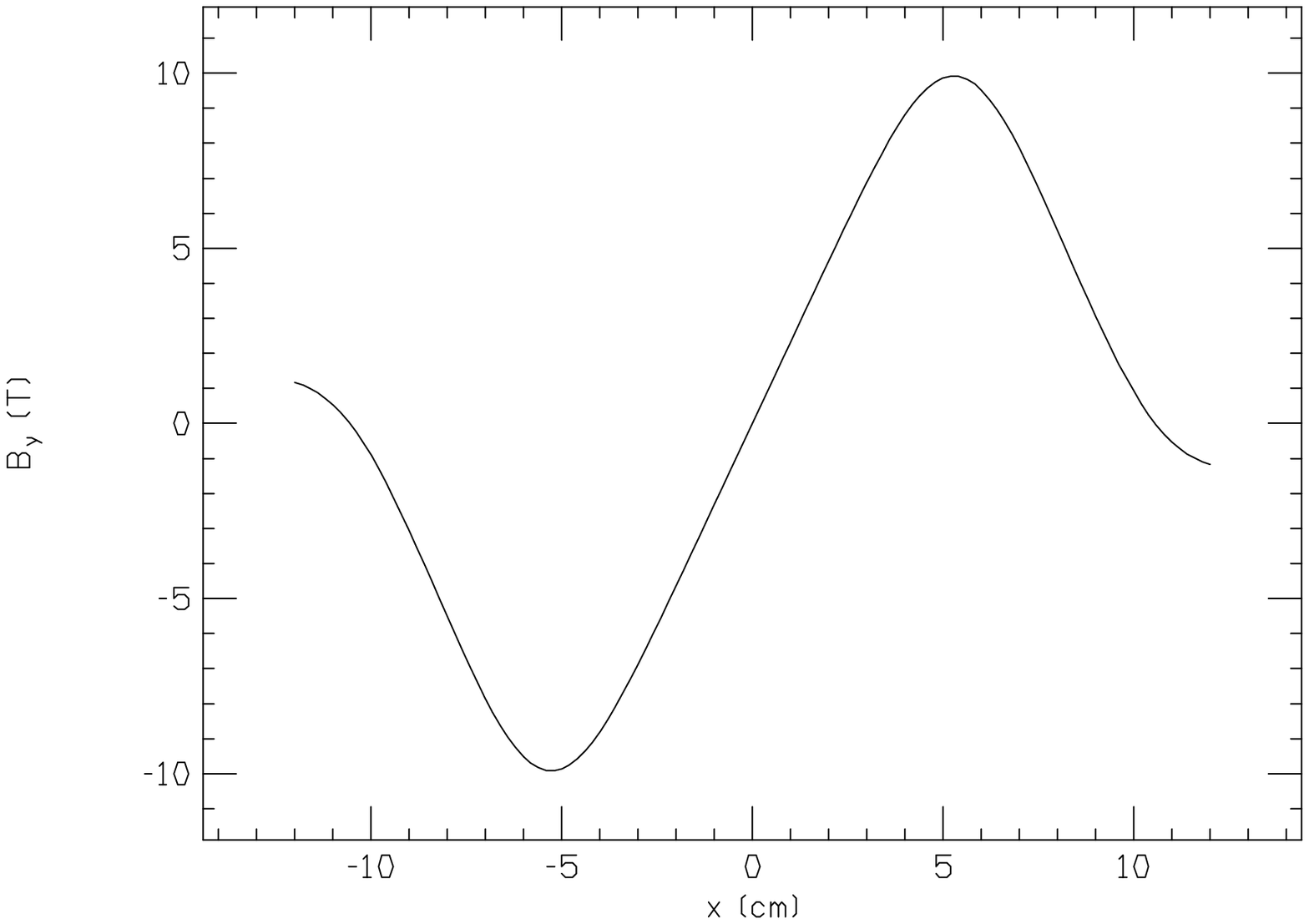,height=4.0in,width=3.5in}}
\caption{Vertical field as function of horizontal position in `C' Quadrupole. }
 \label{cquadB}
 \end{figure}

If this approach were chosen, then the quadrupoles could also be made as `C'
magnets (see Fig.\ref{cquad}). In such magnets there would be two pancake coils 
above and two below the vacuum chamber, with the current directions,  
in the two coils, opposite. This arrangement would generate a downward field to the left of 
the beam and an upward field to the right with a linear gradient, i.e. 
quadrupole field, in the center (see Fig.\ref{cquadB}) Again there appear no 
theoretical problems in defining coil blocks that achieve good field quality, 
and gradients of about 230 T/m seem practical with ${\rm Nb_3 Sn}$ conductors. 
Again, the problems would be in supporting the coils and maintaining the 
required conductor position accuracy. 

\subsubsection{Discussion}
The `C' magnet designs would require about 2/3 of the superconductor needed
for the cos theta magnets, but their overall size would not be much smaller. 
The simple pancake coils of the `C' magnets could be easier to wind than the
cos theta type, but the support of the coils would be a new challenge. It is 
not yet known whether the thermal load would be greater or less in a `C' 
magnet design. Clearly, more study is needed to determine which approach is 
best.

\subsection{Lattice Design}

\subsubsection{Arcs}

In a conventional 2 TeV superconducting ring the tune would be of the order of 
200 and the momentum compaction $\alpha$ around $2 \times 10^{-3}$. In this case, in order to 
maintain a bunch with rms length 3 mm, 45 GeV of S-band rf would be required. 
This would be excessive. It is thus proposed to use an approximately 
isochronous lattice of the dispersion wave type\cite{ref17}. Ideally one would 
like an $\alpha$ of the order of $10^{-7}$. In this case no rf would 
be needed to maintain the bunch and the machine would behave more like a 
linear beam transport. In practice it appears easy to set the zero'th 
order slip factor $\eta_0$ to zero, but if nothing is done, there is a 
relatively large first order slip factor $\eta_1$ yielding a maximum $\alpha$ 
of the order of $10^{-5}$. The use of sextupoles appears able to correct this 
$\eta_1$ yielding a maximum $\alpha$ of the order of $10^{-6}$. With 
octupoles it may be possible to correct $\eta_2$, but this remains to be 
seen. 

It had been feared that amplitude dependent anisochronisity generated in the 
insertion would cause bunch growth in an otherwise purely isochronous design. 
It has, however, been pointed out \cite{oide} that if chromaticity is 
corrected in the ring, then amplitude dependent anisochronisity is 
automatically removed. 

\subsubsection{Low $\beta$ Insertion} 

\begin{figure}[t!] 
\centerline{\epsfig{file=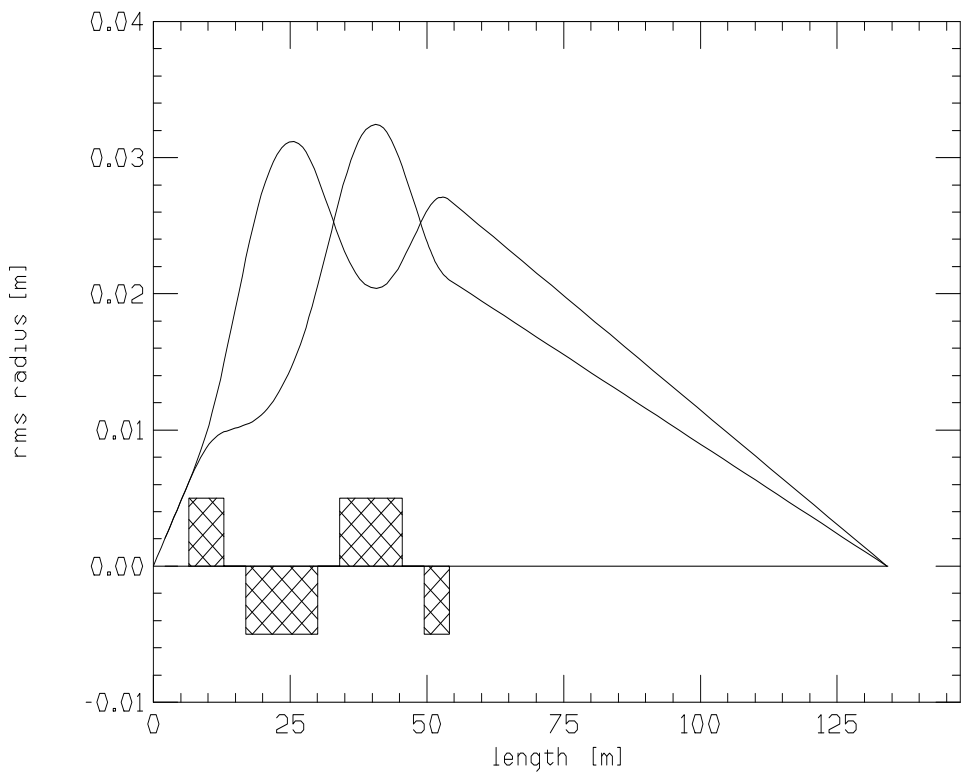,height=4.0in,width=5.0in}}
\caption{{\it rms} radius of the beam at the last four quadrupoles of the 
final focus. }
 \label{ff}
 \end{figure}

In order to 
obtain the desired luminosity we require a very low beta at the intersection 
point: $\beta^*=3\,{\rm mm}$ for 4 TeV, $\beta^*=8\,{\rm mm}$ for the .5 TeV 
design. A possible final focusing quadruplet design is shown in Fig.\ref{ff}. 
The parameters of the quadrupoles for this quadruplet are given in 
Tb.\ref{ffquads}. The maximum fields at 4 sigma have been assumed to be $6.4\,$T. 
This would allow a radiation shield of the order of 5 cm, while keeping the 
peak fields at the conductors less than 10 T, which should be possible using 
${\rm Nb_3Sn}$ conductor. 

With these elements, the maximum beta's in both x and y are of the order of 400 
km in the 4 TeV case, and 14 km in the 0.5 TeV machine. The chromaticities 
$(1/4\pi\int \beta dk)$ are approximately 6000 for the 4 TeV case, and 600 for 
the .5 TeV machine. Such chromaticities are too large to correct within the 
rest of a conventional ring and therefore require local correction\cite{ref18}. 

\begin{table}
 \begin{tabular}{lccccc}
  & &\multicolumn{2}{c}{$4\,$MeV}& \multicolumn{2}{c}{$0.5\,$MeV} \\
\tableline
       & field (T) &  L(m) & R(cm)& L(m) & R(cm)  \\
\tableline
drift  &      & 6.5    &    &   1.99   &         \\
focus  &   6  &  6.43  & 6  &   1.969  & 5.625   \\
drift &       & 4.0    &    &   1.2247 &         \\
defocus & 6.4 & 13.144 & 12 &   4.025  & 11.25  \\
drift &       &  4.0    &    &  1.2247 &         \\
focus  &  6.4 & 11.458 & 12 &   3.508  & 11.25  \\
drift &       &   4.0   &    &  1.2247 &         \\
defocus &6.348& 4.575  & 10 &   1.400  & 9.375  \\
drift   &     &  80    &     & 24.48        &             \\
\end{tabular}
 \caption{Final Focus Quadrupoles; L and R are the length and the radius
respectively. } 
\label{ffquads}
\end{table}

\begin{figure}[bht] 
\centerline{\epsfig{file=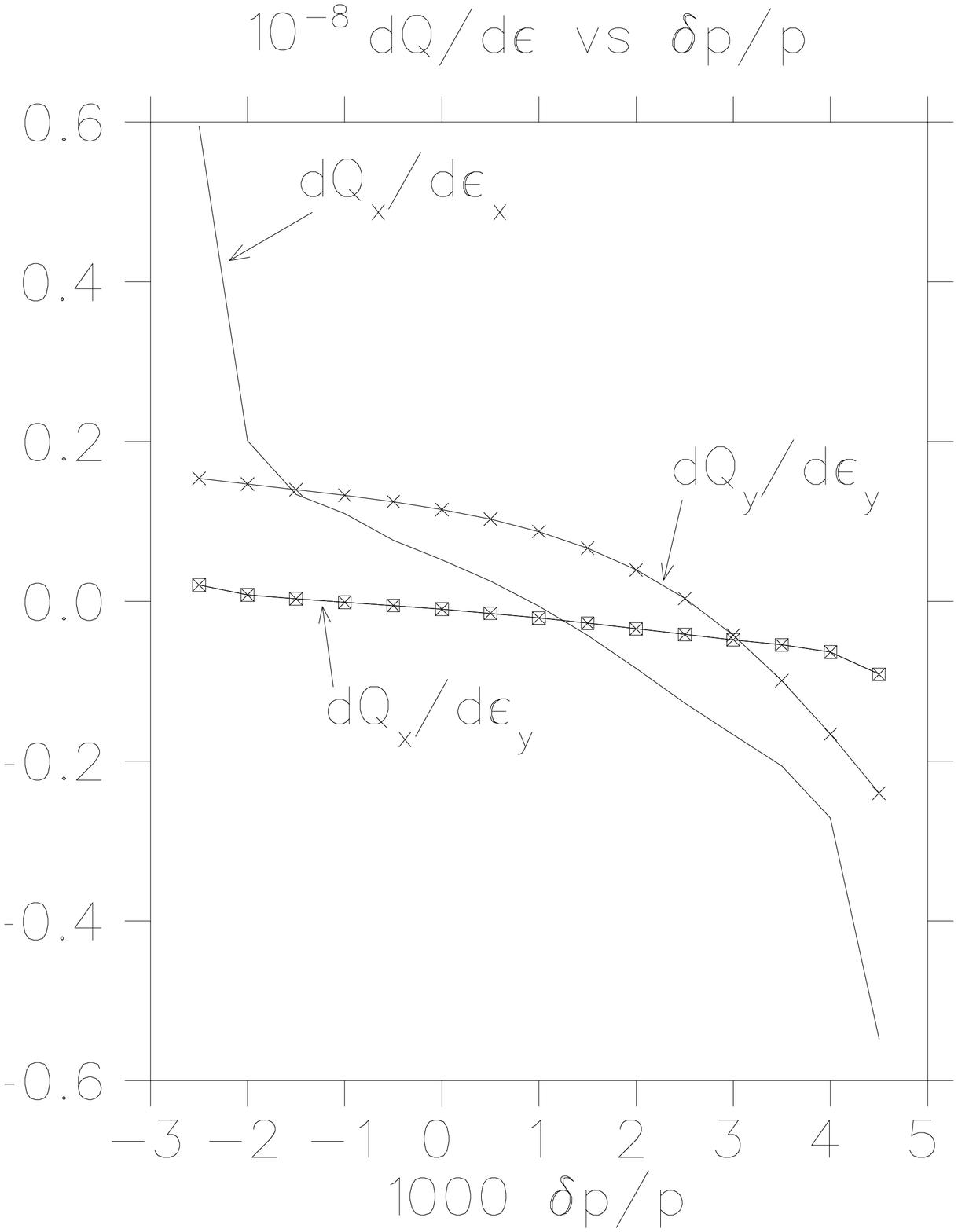,height=4.0in,width=3.5in}}
\caption{Amplitude dependent tune shift ${dQ\over d\epsilon}$ as a 
functions of $\Delta p/p.$  }
\label{ampli}
\end{figure}

   A preliminary {\it model} design\cite{ff} of local chromatic correction 
has been generated for the 4 TeV case, and has been incorporated into a 
dispersion wave lattice\cite{fulllattice}.
%
\begin{figure}[t!] 
\centerline{\epsfig{file=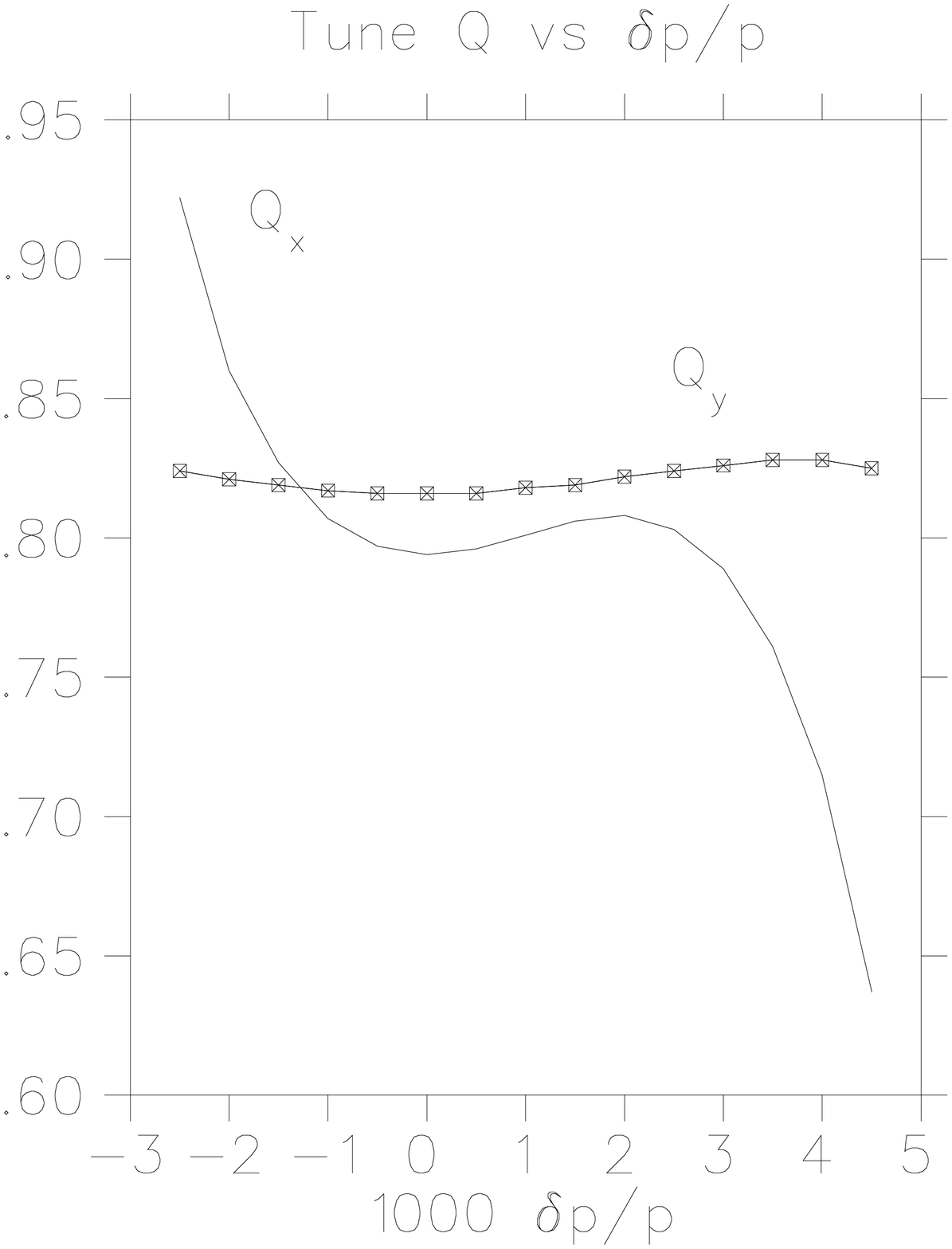,height=4.0in,width=3.5in}}
\caption{The tune shift as a function of $\Delta p/p.$}
 \label{irtune}
 \end{figure}
 Fig.\ref{irtune} shows the tune shift as a function of momentum. It is seen 
that this design has a momentum acceptance of $\pm 0.35\,\%$. The second order 
amplitude dependent tune shifts shown in Fig.\ref{ampli} are less than 0.03 at one sigma (0.27 at 3 
sigma in amplitude), which may be too large, even for only 1000 turns. In 
addition, this design used some bending fields that are unrealistic. It is 
expected that these limitations will soon be overcome, and that more 
sophisticated designs\cite{ref19} should achieve a momentum acceptance of $\pm 
0.6\,\%$ for use with a clipped rms momentum spread of 0.2 \%. 
\subsection{Instabilities}
Studies\cite{stability} of the  resistive wall impedance instabilities indicate 
that the required muon bunches (eg for $2\,$TeV: $\sigma_z=3\ mm,\ N_{\mu}= 
2\times 10^{12}$) would be unstable in a conventional ring. In any case, the 
rf requirements to maintain such bunches would be excessive. 

If one can obtain momentum-compaction factor  $\alpha \leq 10^{-7}$, then the 
synchrotron oscillation period is longer than the effective storage time, and 
the beam dynamics in the collider behave like that in a linear beam 
transport\cite{ngstab}\cite{chengstab}. In this case, beam breakup 
instabilities are the most important collective effects.  Even with an aluminum
beam pipe of  radius  $b=2.5$~cm, the resistive wall effect will cause the tail
amplitude of  the bunch to double in about 500~turns.  For a broad-band
impedance of $Q=1$  and $Z_\parallel/n=1$~Ohm, the doubling time in the same
beam pipe is only  about 130 turns; which is clearly unacceptable. But both
these instabilities can easily be  stabilized using BNS\cite{bns} damping. For
instance, to stabilize the resistive wall instability, the required tune
spread, calculated\cite{ngstab} using the two particle
model approximation, is (for Al pipe)
\begin{equation}
{\Delta \nu_{\beta} \over \nu_{\beta}}=\left\{ \begin{array}{ll}
1.58\,10^{-4} & b=1.0\,{\rm cm}\\
1.07\,10^{-5} & b=2.5\,{\rm cm}\\ 
1.26\,10^{-6} & b=5.0\,{\rm cm}
\end{array}
\right.
\end{equation} 
 But  this application of the BNS damping to a quasi-isochronous ring, where
there  are other head-tail instabilities due to the  chromaticities $\xi$ and $
\eta_1$, needs more careful study. 

If it is not possible to obtain an $\alpha$ less than $10^{-7}$, then 
rf must be introduced and synchrotron oscillations will occur. The above 
instabilities are then somewhat stabilized because of the interchanging of head 
and tail, but the impedance of the rf now adds to the problem and simple 
BNS damping is no longer possible. 

For example, a momentum-compaction  factor $|\alpha|\approx1.5\times10^{-5}$
has been studied; rf of $\sim 1.5$~GV is  needed which gives a synchrotron
oscillation period of 150 turns.  Three  different impedance models: 
resonator,  resistive wall, and a  SLAC-like or a  CEBAF-like rf accelerating
structure have been used in the estimation for  three sets of design
parameters. The impedance of the ring is dominated by  the rf cavities, and the
microwave instability is well beyond threshold. Two  approaches are being
considered to control these instabilities: 1) BNS damping  applied by rf
quadrupoles as suggested  by Chao\cite{chaobook}; and 2) applying an 
oscillating perturbation on the chromaticity\cite{cheng2}. 

When the ring is nearly isochronous, a longitudinal head-tail (LHT) instability 
may occur because the nonlinear slip factor $\eta_1$ becomes more important 
than the first order $\eta_0$ [3]. The growth time for the rf impedance when 
$\eta \simeq 10^{-5}$ is about $0.125 b \eta_0/\eta_1$~s, where $b$ is the 
pipe radius in cm.  This would be longer than the storage time of $\sim 41$~ms 
if $\eta_1 \sim \eta_0$. However, if $\eta_1 \sim \eta_0/\delta$, with 
$\delta \sim 10^{-3}$, then the growth time is about $0.125 b $~ms, which 
is much shorter than the storage time. More study is needed.

\section{COLLIDER PERFORMANCE}
\subsection{Luminosity vs Energy and Momentum Spread}
  The bunch populations decay exponentially, yielding an integrated luminosity 
equal to its initial value multiplied by an {\it effective} number of turns 
$n_{{\rm effective}}=150\ B,$ where B is the mean bending field in T. 

The luminosity is given by:
 \b
\Ls\={N^2\ f\ n_e \gamma\over 4\pi\ \beta^*\ \epsilon_n} H(A,D)\label{lumi}
 \e
where $A=\sigma_z / \beta^*$,
 \b
D={\sigma_z N\over \gamma \sigma_t^2} r_e ({m_e\over m_{\mu}})
 \e
and the enhancement factor is 
 \b
H(A,D)\approx 1+D^{1/4} \left[ {D^3\over 1+D^3} \right] \left\{
\ln{(\sqrt{D}+1)} + 2\ln{({0.8\over A})} \right\}.
 \e
In our case A = 1, D $\approx$ .5 and H(A,D) $\approx 1$\cite{pisin2}.
  
For a fixed collider lattice, operating at energies lower than the design 
value, the luminosity will fall as $\gamma^3.$ One power comes from the 
$\gamma$ in Eq.\ref{lumi}; a second comes from $n_e$, the effective 
number of turns, that is proportional to ${\gamma}$; the third factor comes from 
$\beta^*,$ which must be increased proportional to $\gamma$ in order to keep 
the beam size constant within the focusing magnets. The bunch length 
$\sigma_z$ must also be increased proportional to $\gamma$ so that the 
required longitudinal phase space is not decreased; so A = $\sigma_z/\beta^*$ 
remains constant. 

In view of this rapid drop in luminosity with energy, it would be desirable
to have separate collider rings at relatively close energy spacings: e.g.
not more than factors of two apart.

   If it is required to lower the energy spread $\Delta E/E$ at a fixed
energy, then again the luminosity will fall.  Given the same longitudinal phase space, the bunch length
$\sigma_z$ must be increased. If the final focus is retuned to simultaneously
increase $\beta^*$ to maintain the same value of $A,$ then the luminosity will be
exactly proportional to  $\Delta E/E.$ But if, instead, the $\beta^*$ is kept
constant, and the parameter A allowed to increase, then the luminosity falls
initially at a somewhat lower rate. The luminosity, for small  $\Delta E/E$ is
then approximately given by:
 \b
\Ls\=2\ \Ls_0\ {\Delta E/E\over {(\Delta E/E)}_0}.
 \e
There may, however, be beam-beam tune shift emittance growth problems in this case.
\subsection{Detector Background}
There will be backgrounds from the decay of muons in the ring, from muon halo
around the beam, and from the interactions themselves.

\subsubsection{Muon Decay Background}
  A first Monte Carlo study\cite{ref20} of the muon decay background was done 
with the MARS95 code\cite{MARS}, based on a preliminary insertion lattice. A 
tungsten shielding {\it nose} was introduced, extending to within 15 cm of the 
intersection point. It was found that: 
\begin{itemize} 
   \item{a large part of the electromagnetic background came from synchrotron
radiation, from the bending magnets in the chromatic correction
section.
   \item as many as 500 hits per cm$^2$ were expected in a vertex detector,
falling off to the order of 2 hits per cm$^2$ in an outer tracker. }
   \item{there was considerable, very low energy, neutron background: of the
order of 30,000 neutrons per cm$^2,$ giving, with an efficiency of $3\ 10^{-
3}$, about 100 hits per cm$^2$. }
  \end{itemize}

   It was hoped that by improving the shielding these backgrounds could be
substantially reduced.

   A more recent study\cite{iuliu} of the electromagnetic component of the
background has been done using the GEANT codes\cite{geant}. This study
differed from the first in several ways:
   \begin{itemize}
   \item{ the shower electrons and photons were followed down to a
lower energy ($50\,$keV for electrons and $15\,$keV for photons).}
   \item{the {\it nose} angle, i.e. the angle not seen by the detector, was
increased
from 9 to 20 degrees to reduce radial shower penetration.}
   \item{the {\it nose} design was modified (see Fig.\ref{nose}) so that: 1) 
The incoming electrons are collimated to $\pm\ 4\ \times\ \sigma_{\theta_0}$ 
(where $\sigma_{\theta_0}$ is the rms divergence of the beam) by a cone 
leading down towards the vertex. 2)  The detector could not {\it see} any 
surface directly illuminated by these initial electrons, whether seen in the 
forward or backward (albedo) directions. 3)  The detector could not {\it see } 
any surface that is illuminated by secondary electrons if the secondary 
scattering angle is forward. 4)  The minimum distance through the collimator 
from the detector to any primarily illuminated surface was more than 100 mm, 
and from any secondarily illuminated surface, 30 mm.                      } 
 \begin{figure}[t!] 
\centerline{\epsfig{file=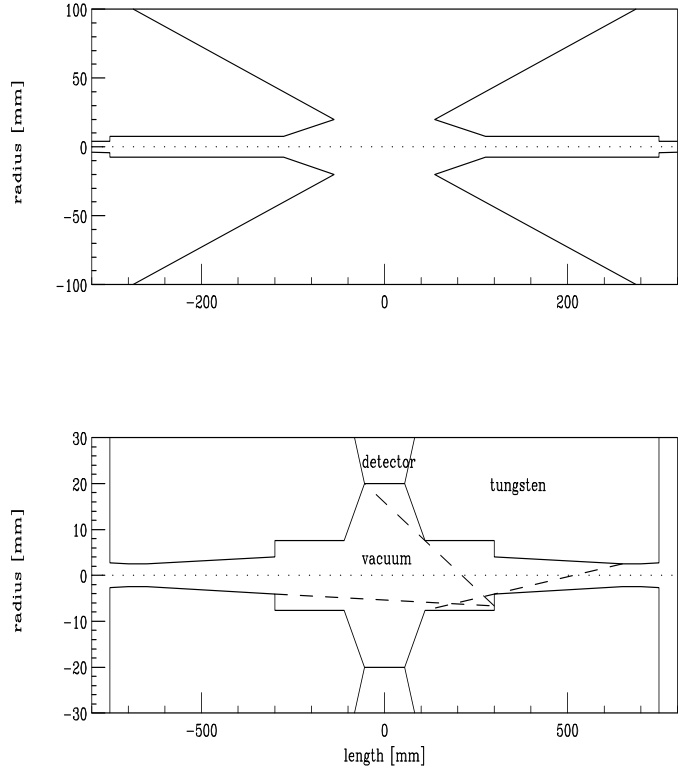,height=4.0in,width=3.5in}}
\caption{Schematic of the detector nose.}
 \label{nose}
 \end{figure}
   \item{it was assumed that a collimator placed at a focus  $130\,{\rm m}$ from the 
intersection point would be able to effectively shield all synchrotron photons 
from the bending magnets beyond that point. The {\it rms} beam size at this 
focus is only 10 $\mu m$ so a very effective collimation should be possible
(see Fig.\ref{ff}). } 
   \end{itemize}

This study indicated that the dominant background was no longer from synchrotron
photons, but from photons from $\mu$ decay electrons. The average momentum of these photons
was only 1 MeV.  Tb.\ref{background} gives the total numbers of photons,
the total number of hits, possible pixel sizes, and the hits per pixel, for a)
a vertex detector placed at a 5 cm radius, and b) a gas detector placed at a 1
m radius. In all cases the numbers are given per bunch crossing.
\begin{table}
\begin{tabular}{lcc}
Detector              &   vertex     &   tracker  \\
Radius                &      5 cm      &    1 m   \\
\tableline
Number of photons     & $50\ 10^6$   & $15\ 10^6$    \\
Number of hits        &   150,000     &  15,000      \\
Detector Area         &   863 cm$^2$  &  34 m$^2$   \\
Pixel size            & 20 x 20 $\mu$m& 1 mm x 1 cm    \\
Sensitivity           &   0.3 \%     &    0.1 \%      \\
Occupancy             & .07 \%       &    0.4 \%     \\
\end{tabular}
\caption{Detector Backgrounds from $\mu$ decay}
\label{background}
\end{table}
   The sensitivities given here are for a silicon strip detector (.3\%)
at the small 
radius, and a pad readout gas chamber (.1\%)
at the larger radius. Both sensitivities 
could be reduced. Silicon strip detectors could be developed with less 
thickness, and a time projection chamber (TPC) could be used at the larger radii. 
The use of the TPC would be particularly advantageous because, not only is its
density lower, but the small depositions of ionization from low energy 
photons and neutrons could not be mistaken for real tracks. 

   This study also found a relatively modest flux of muons from $\mu$ pair production
in electromagnetic showers: about 50 such tracks pass through the detector per
bunch crossing.

   The general conclusion of the two studies are not inconsistent as a cursory
look may indicate. The 
background, though serious, is not impossible to overcome. Further reductions 
are expected as the shielding is optimized, and, as mentioned above, it should 
be possible to design detectors that are less sensitive to the neutrons and 
photons present. 

\subsubsection{Muon Halo Background}
   There would be a very serious background from the presence of even a very
small halo of nearly full energy muons in the circulating beam\cite{strig}. The beam will
need careful preparation before injection into the collider, and a collimation
system will have to be designed to be located on the opposite side of the ring
from the detector.
\subsubsection{Electron Pair Background}

   In \ee machines there is a significant problem from beamstrahlung photons
(synchrotron radiation from beam particles in the coherent field of the oncoming
bunch), and an additional problem from pair production by these photons.

   With muons, there is negligible beamstrahlung, and thus negligible pair 
production from them. P. Chen\cite{beamstrahlung} has further 
shown that beamstrahlung of electrons from the nearby decay of muons does not 
pose a problem. 

   There is, however, significant incoherent (i.e. \mumu $\rightarrow$ \ee) 
pair production in the 4 TeV Collider case. The cross section is estimated to 
be $10\,$mb\cite{pairsection}, which would give rise to a background of 
$\approx 3\,10^4$  electron pairs per bunch crossing. Approximately 
$90\,\%$ of these, will be trapped inside the tungsten nose cone, but 
those with energy between 30 and $100\,$MeV will enter the detector region. 

   There remains some question about the coherent pair production generated by 
the virtual photons interacting with  the coherent electromagnetic fields of 
the entire oncoming bunch. A simple Weizs\"acker-Williams 
calculation\cite{pisin} yields a background that  would 
consume the entire beam at a rate comparable with its decay. However, I. 
Ginzburg\cite{ginzburg} and others have argued that the integration must be 
cut off due to the finite size of the final electrons. If this is true, then 
the background becomes negligible.  A more detailed study of this problem is 
now underway\cite{pisinandkroll}. 

   If the coherent pair production problem is confirmed, then there are two 
possible solutions: 

   1) one could design a two ring, four beam machine (a $\mu^+$ and a
$\mu^-$ bunch coming from each side of the collision region, at the same
time). In this case the coherent electromagnetic fields at the intersection
are canceled and the pair production becomes negligible.

   2) plasma could be introduced at the intersection point to cancel the beam
electromagnetic fields\cite{plasma}.

\subsection{Polarization}

\subsubsection{Polarized Muon Production}
   The generation of polarized muons has not yet received enough attention and 
the specifications and components described above have not been designed or 
optimized for polarization. Nevertheless, simple manipulations of parameters 
and/or the addition of simple components would allow some polarization with 
relatively modest loss of luminosity. 

   In the center of mass of a decaying pion, the outgoing muon is fully 
polarized (-1 for $\mu^+$ and +1 for $\mu^-$). In the lab system the 
polarization depends\cite{decaypolar} on the decay angle $\theta_d$ and 
initial pion energy. For pion kinetic energy larger than the pion mass, the 
dependence on pion energy becomes negligible and the polarization is given 
approximately by:
 \b
P_{\mu^-} \approx \cos{\theta_d} + 0.28(1 -\cos^2{\theta_d})
 \e
The average value of this is about 0.19. At lower pion energies the 
polarization is higher, and has a value of the order of 0.5 at a kinetic 
energy of 10 MeV. If nothing is done, the polarization of the muons captured 
and phase rotated by the proposed system is approximately 20 \%. 

If higher polarization is required, some selection of muons from forward pion
decays  $(\cos{\theta_d} \rightarrow 1)$ is required. This could be done by 
selecting pions within a narrow energy range and then selecting only those 
muons with energy close to that of the selected pions. But such a procedure 
would collect a very small fraction of all possible muons and would yield a 
very small luminosity. Instead we wish, as in the unpolarized case, to capture 
pions over a wide energy range, allow them to decay, and to use rf to phase 
rotate the resulting distribution. 

\begin{figure}[bht] 
\centerline{\epsfig{file=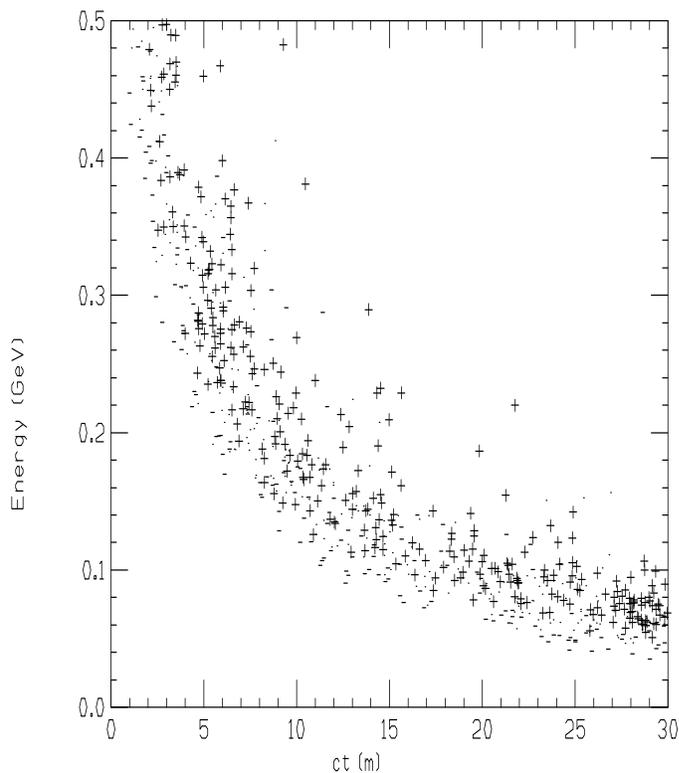,height=4.0in,width=3.5in}}
\caption{Energy vs ct of $\mu$'s at end of decay channel ( no phase 
rotation).}
 \label{Evsctpol1}
 \end{figure}

Consider the distributions in velocity vs ct at the end of a decay  channel.
If the source bunch of protons is very short and if the pions were  generated
in the forward direction, then the pions, if they did not decay, would all be
found  on a single curved line. Muons from forward decays would have gained
velocity  and would lie above that line. Muons from backward decays would have
lost  velocity and would fall below the line. The real distribution would be
diluted  by the width of the proton bunch and the finite pion angles. In order
to  reduce the latter, it is found desirable to lower  the solenoid field in
the  decay channel from 5 to 3 Tesla. When this is done one obtains the 
distribution shown in Fig.\ref{Evsctpol1}, where the polarization P$>{1\over 3}$,
${-1\over 3}< P<{1\over 3}$, and P$<{-1\over 3}$ is marked by the symbols
`+', `.' and `-' respectively. 

\begin{figure}[hbt] 
\centerline{\epsfig{file=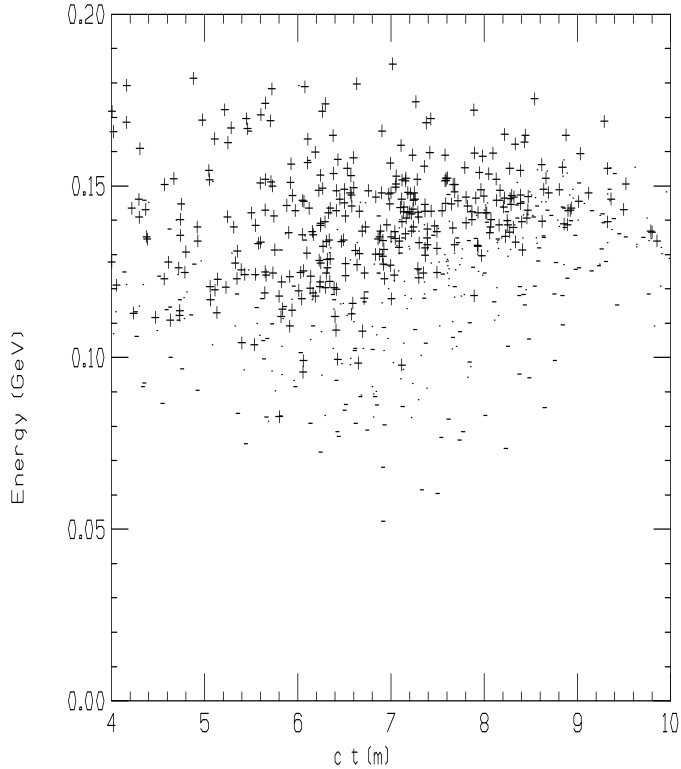,height=4.0in,width=3.5in}}
\caption{Energy vs ct of $\mu$'s at end of decay channel with phase 
rotation.}
 \label{Evsctpol2}
 \end{figure}

After phase rotation with rf the 
correlation is preserved: see Fig.\ref{Evsctpol2} where as before the polarization P$>{1\over 3}$, ${-1\over 3}< P<{1\over 3}$, 
and P$<{-1\over 3}$ is marked by the symbols `+', `.' and `-' respectively.
 
\begin{figure}[bht] 
\centerline{\epsfig{file=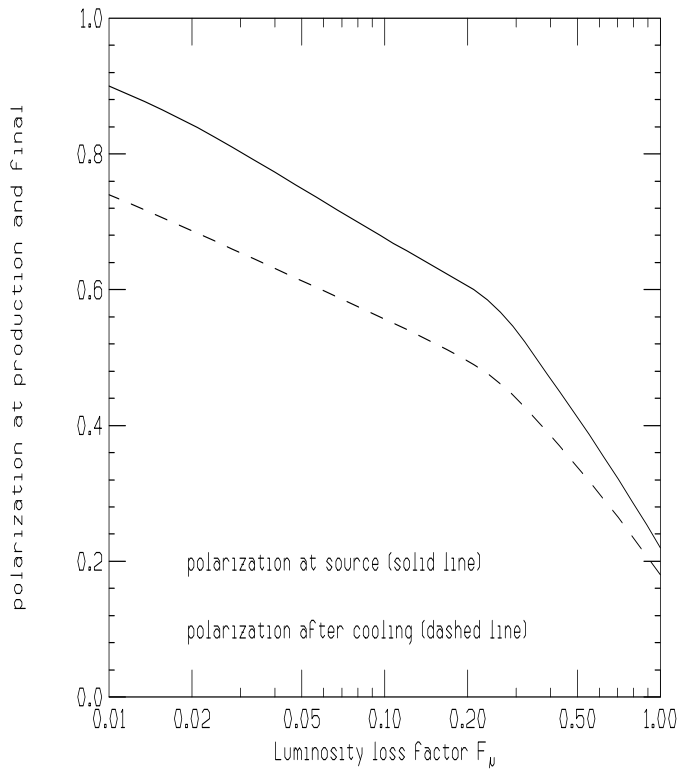,height=4.0in,width=3.5in}}
\caption{Polarization vs $F_{\mu}$ of $\mu$'s accepted.} 
 \label{polvscut}
 \end{figure}

   If a selection is made on the minimum energy of the muons, then net 
polarization is obtained. The tighter the cut on energy, the higher the 
polarization, but the less the fraction $F_{\mu}$ of muons that are selected. 
Fig.\ref{polvscut} gives the results of a Monte Carlo study.

The loss, about  30\%, from the use of the lower solenoid field, is included in
the fractions  $F_{\mu}$ plotted. 

%
\subsubsection{Polarization Preservation}
   A recent paper\cite{rossmanith} has discussed the preservation of muon
polarization in some detail. 
During the ionization cooling process the muons lose energy in material and 
have a spin flip probability ${\cal P},$ 
 \b
{\cal P}\approx \int {{m_e}\over {m_{\mu}}}\beta_v^2\ {dE\over E }
 \e
where $\beta_v$ is the muon velocity divided by c, and dE/E is the fractional 
loss of energy due to ionization loss. In our case the integrated energy loss 
is approximately 3 GeV and the typical energy is 150 MeV, so the integrated 
spin flip probability is close to 10\%. The change in polarization $dP/P$ is 
twice the spin flip probability, so the reduction in polarization is 
approximately 20~\%. 

   During circulation in any ring, the muon spins, if initially longitudinal, 
will precess by (g-2)/2 $\gamma$ turns per revolution in the ring; where
(g-2)/2 is $1.166\ 10^{-3}$. A given energy spread $d\gamma/\gamma$ will 
introduce variations in these precessions and cause dilution of the polarization. 
But if the particles remain in the ring for an exact integer number of synchrotron 
oscillations, then their individual average $\gamma$'s will be the same and no 
dilution will occur. It appears reasonable to use this `synchrotron spin 
matching'\cite{rossmanith} to avoid dilution during acceleration.

   In the collider, however, the synchrotron frequency will probably be too 
slow to use `synchrotron spin matching', so one of two methods must be used.

\begin{itemize}
 \item{ Bending can be performed with the spin orientation in the vertical 
direction, and the spin rotated into 
the longitudinal direction only for the interaction region. The design of such 
spin rotators appears relatively straightforward. The example given in the 
above reference would only add 120 m of additional arc length, but no design 
has yet been incorporated into the lattice.} 
 \item{ The alternative is to install a 120 m 10 T solenoid (Siberian Snake) 
at a location exactly opposite to the intersection point.  Such a solenoid 
reverses the sign of the horizontal polarization and generates a cancellation 
of the precession in the two halves of the ring.} 
 \end{itemize} 

   Provision must also be made to allow changes in the relative spins of the 
two opposing  bunches. This could be done, prior to acceleration, by switching 
one of the two beams into one or the other of two alternative injection lines. 

\subsubsection{Luminosity loss with polarization}

   If both muon beams are polarized, then naturally the luminosity would drop 
as the square of the fraction $F_{\mu}$ of selected $\mu$'s shown in 
Fig.\ref{polvscut}, but the loss need not be so great. In the unpolarized case 
of the $4\,$TeV collider, there were two bunches of each sign. If the 
$F_{\mu}$  chosen for polarization is less than 0.5, then this number of 
bunches could be reduced to one, without introducing excessive beam-beam tune 
shift. A factor of two in luminosity is then restored. If, for instance, the 
$F_{\mu}$ is taken as 0.5 for both signs, and the number of bunches is 
reduced to one of each sign, then the luminosity is reduced by a factor of 
only 0.5 and not 0.25. 

   One also notes that the luminosity could be maintained at the full 
unpolarized value if the proton source intensity could be increased. Such an 
increase in proton source intensity in the unpolarized case would be 
impractical because of the resultant excessive high energy muon beam power, 
but this restriction does not apply if the increase is used to offset losses 
in generating polarization. If, for instance, the driver repetition rate were 
increased from 15 to 30 Hz, the fractions $F_{\mu}$   set at 0.5, and the 
number of bunches reduced to one, then the full luminosity of $10^{35}\ (cm^{-
2} s^{-1})$ would be maintained with polarization of both beams of 35\%. 

   The numbers given in this section are preliminary. Optimization of 
the systems may improve the polarizations obtained, but other dilution 
mechanisms may reduce them. 

\section{CONCLUSION}
 \begin{itemize}
 \item Considerable progress has been made on a scenario for a 2 + 2 TeV, high 
luminosity collider. Much work remains to be done, but no obvious show stopper 
has yet been found. 
 \item The two areas that could present serious problems are: 1) unforeseen 
losses during the 25 stages of cooling (a 3\% loss per stage would be very 
serious); and 2) the excessive detector background from muon beam halo. 
 \item Many technical components require development: a large high field
solenoid for capture, low frequency rf linacs, multi-beam
pulsed and/or rotating magnets for acceleration, warm bore shielding inside
high field dipoles for the collider, muon collimators and background shields,
etc.\ but:
 \item None of the required components may be described as {\it exotic}, 
and their specifications are not far beyond what has been demonstrated.
 \item If the components can be developed and the problems can be overcome,
then a muon-muon collider should be a viable tool for the study of high energy
phenomena, complementary to \ee and hadron colliders.
 \end{itemize}

\section{ACKNOWLEDGMENTS}
We acknowledge important contributions from our colleagues, especially 
W. Barletta, A. Chao, J. Irwin, H. Padamsee, C. Pellegrini and A. Ruggiero.  

\medskip
This research was supported by the U.S. Department of Energy under Contract No.
DE-ACO2-76-CH00016 and DE-AC03-76SF00515.


\end{document}